\documentclass[a4paper,11pt]{article}
\pdfoutput=1 % if your are submitting a pdflatex (i.e. if you have
\usepackage{jcappub} % for details on the use of the package, please
\usepackage[T1]{fontenc} % if needed

\usepackage{amsmath}
\usepackage{graphicx,epsf,hyperref}

\usepackage{color}
\usepackage[dvipsnames]{xcolor}

\newcommand{\Hh}{\mathcal{H}}
\newcommand{\n}{\nabla}
\newcommand{\p}{\partial}

\newcommand{\bfk}{\mbox{\boldmath$k$}}

\newcommand{\bfq}{\mbox{\boldmath$q$}}

\newcommand{\bfn}{\mbox{\boldmath$n$}}
\newcommand{\de}{\mathrm{d}}

\def\be{\begin{equation}}
\def\ee{\end{equation}}

\title{Interacting dark energy from redshift-space galaxy clustering}

\author{Pedro Carrilho$^a$, }

\author{Chiara Moretti$^{a,b}$, }

\author{Benjamin Bose$^c$, }

\author{Katarina Markovič$^d$, and}

\author{Alkistis Pourtsidou$^{a,e}$}

\affiliation{$^a$ Astronomy Unit, School of Physics and Astronomy, Queen Mary University of London,\\
Mile End Road, London, E1 4NS, UK}

\affiliation{$^b$ INAF – Osservatorio Astronomico di Trieste, Via Tiepolo 11, I-34143 - Trieste, Italy}

\affiliation{$^c$ Département de Physique Théorique, Université de Genève, 24 quai Ernest Ansermet,
1211 Genève 4, Switzerland}

\affiliation{$^d$ Jet Propulsion Laboratory, California Institute of Technology, 4800 Oak Grove Drive, Pasadena, CA, USA}

\affiliation{$^e$ Department of Physics \& Astronomy, University of the Western Cape, Cape Town 7535, South Africa}

\emailAdd{p.gregoriocarrilho@qmul.ac.uk}
\emailAdd{c.moretti@qmul.ac.uk}

\abstract{
Interacting dark energy models have been proposed as attractive alternatives to $\Lambda$CDM.
Forthcoming Stage-IV galaxy clustering surveys will constrain these models, but they require accurate modelling of the galaxy power spectrum multipoles on mildly non-linear scales. In this work we consider a dark scattering model with a simple 1-parameter extension to $w$CDM -- adding only $A$, which describes a pure momentum exchange between dark energy and dark matter. We then provide a comprehensive comparison of three approaches of modeling non-linearities, while including the effects of this dark sector coupling. We base our modeling of non-linearities on the two most popular perturbation theory approaches: TNS and EFTofLSS. To test the validity and precision of the modelling, we perform an MCMC analysis using simulated data corresponding to a $\Lambda$CDM fiducial cosmology and Stage-IV surveys specifications in two redshift bins, $z=0.5$ and $z=1$. We find the most complex EFTofLSS-based model studied to be better suited at both, describing the mock data up to smaller scales, and extracting the most information. Using this model, we forecast uncertainties on the dark energy equation of state, $w$, and on the interaction parameter, $A$, finding $\sigma_w=0.06$ and $\sigma_A=1.1$ b/GeV for the analysis at $z=0.5$ and $\sigma_w=0.06$ and $\sigma_A=2.0$ b/GeV for the analysis at $z=1$. In addition, we show that a false detection of exotic dark energy up to 3$\sigma$ would occur should the non-linear modelling be incorrect, demonstrating the importance of the validation stage for accurate interpretation of measurements. 

}

\begin{document}

\maketitle

\section{Introduction}

Current and forthcoming large-scale structure (LSS) cosmological experiments such as the Dark Energy Survey (DES)\footnote{\url{https://www.darkenergysurvey.org/}}~\cite{Abbott:2021bzy}, the Dark Energy Spectroscopic Instrument (DESI)\footnote{\url{https://www.desi.lbl.gov/}}~\cite{Aghamousa:2016zmz}, 
Euclid\footnote{\url{http://euclid-ec.org}}~\cite{Blanchard:2019oqi,Laureijs:2011gra}, the Nancy Grace Roman Space Telescope\footnote{\url{https://www.nasa.gov/roman}}~\cite{spergel2015wide}, and the Vera C. Rubin Observatory’s Legacy Survey of Space and Time (LSST)\footnote{\url{https://www.lsst.org/}}~\cite{Mandelbaum:2018ouv} are promising to deliver high precision cosmological measurements, allowing to probe the nature of dark energy. 
The standard model of cosmology, $\Lambda$CDM, has enjoyed immense success, having shown great consistency with a wealth of cosmic microwave background (CMB) and LSS data \cite{Aghanim:2018eyx, Anderson_2012, Song:2015oza, Beutler_2016, SpurioMancini:2019rxy, Troster:2019ean, Alam:2020sor, Abbott:2021bzy,Troster:2020kai}. The model assumes that General Relativity (GR) is the correct description of gravity on all scales, and that dark matter and dark energy are uncoupled. Alongside the standard cosmological model, a plethora of alternative models have been proposed. These include dark energy in the form of quintessence, modified gravity, and coupled models of dark matter and dark energy \cite{Caldwell:1997ii,Amendola:1999er,Peebles:2002gy,Copeland:2006wr,Nojiri:2006ri,Sotiriou:2008rp, DeFelice:2010aj,Clifton:2011jh,Bertolami:2012xn, Pourtsidou:2013nha}.   

Stage IV spectroscopic galaxy redshift surveys such as the Euclid satellite mission aim to measure the logarithmic growth rate of structure $f$, which strongly depends on cosmology and gravity \cite{Guzzo:2008ac}. This will be done by probing the redshift space distortion (RSD) signature with summary statistics, such as the galaxy power spectrum or correlation function \cite{Blake:2011rj, Reid:2012sw, Beutler:2013yhm, Macaulay_2013,Gil-Marin:2016wya, Blanchard:2019oqi}. With the volume and number density of galaxies probed being large, observations will no longer be limited by statistical errors, but by theoretical, sky, and instrumental systematic uncertainties. The most important theoretical systematic is related to our ability to model the non-linear scales of structure formation, and get precise and unbiased constraints on cosmological parameters. This affects both main LSS probes, i.e., galaxy clustering and weak gravitational lensing \cite{Chisari:2019tus, Markovic:2019sva, Bose:2019psj, Schneider:2019snl, Nishimichi:2020tvu, Martinelli:2020yto, Pezzotta:2021vfn, Secco:2021vhm}. 

In the context of $\Lambda$CDM and redshift-space galaxy clustering, the non-linear modelling challenge has been most widely addressed by two competing perturbative models, the TNS model \cite{Taruya:2010mx} and the effective field theory of large scale structure (EFTofLSS) \cite{Baumann:2010tm,Carrasco:2012cv}. Recent analyses of the BOSS DR12 power spectrum multipoles \cite{Beutler_2016} utilise both of these approaches and find consistent results  \cite{DAmico:2019fhj, Ivanov:2019pdj, Troster:2019ean}.

In order to constrain non-standard cosmologies with forthcoming data, we need to perform extensive non-linear modelling studies and validation tests against simulations. This paper aims to do this for a phenomenologically interesting class of models that includes a non-gravitational interaction (coupling) between dark energy and cold dark matter. Interacting dark energy models have been gaining popularity recently due to their ability to alleviate some cosmological tensions~\cite{2020arXiv200811284D,DiValentino:2020vvd,2021APh...13102607D}. Recent local universe observations have shown mild to large inconsistencies with the CMB observations (see \cite{Verde:2019ivm,Knox:2019rjx,Jedamzik_2021,Di_Valentino_2021,perivolaropoulos2021challenges} for recent reviews). In particular, discrepancies between the Planck CMB measurements of $\Lambda$CDM parameters \cite{Aghanim:2018eyx} and late time galaxy observations \cite{Abbott:2020knk,2021A&A...646A.140H,Vikhlinin:2008ym,Reid:2012sw,Gil-Marin:2016wya,deHaan:2016qvy,Blake:2011rj,Beutler:2013yhm,Simpson:2015yfa} have been uncovered. This tension lies in the growth of structure, specifically the amplitude of density fluctuations, typically characterised by the density fluctuations averaged over a sphere of $8 \, h/{\rm Mpc}$ radius, $\sigma_8$. Late time measurements consistently prefer less growth of structure than the prediction from a $\Lambda$CDM model fit to the CMB. 

One promising means of alleviating the $\sigma_8$ tension is to allow for an interaction between dark matter and dark energy. The most common interacting dark energy models, which assume energy transfer at the background and linear perturbations level, fail to fit CMB data and are severely constrained by data \cite{Bean_2008,Xia:2009zzb,Amendola_2012,Gomez-Valent:2020mqn}, except when the interaction is proportional to the dark energy density~\cite{Di_Valentino_2021}. However, the class of models we study in this work are uncoupled in the background and only exhibit momentum exchange at the level of linear perturbations, allowing them to fit both CMB and LSS data very well \cite{Pourtsidou:2016ico}. 
These models have been formally described and parameterised in \cite{Pourtsidou:2013nha,Skordis:2015yra,Richarte:2014yva} using a Lagrangian approach, and a phenomenological model called the `Dark Scattering' model \cite{Simpson:2010vh} has been incorporated in $N$-body codes \cite{Baldi:2016zom}. Other relevant works that have studied pure momentum exchange models and their ability to alleviate the $\sigma_8$ tension are \cite{Lesgourgues:2015wza,Linton:2017ged,Bose:2017jjx,Buen-Abad:2017gxg,Kase:2019mox,Chamings:2019kcl,Amendola:2020ldb,1869592}.

In this work, we test the capability of three different perturbative models of constraining momentum exchange in the dark sector by stage IV galaxy surveys. We work in the context of the Dark Scattering model, allowing the growth of structure to be affected by the transfer of momentum. We make use of a set of $\Lambda$CDM PICOLA simulations with which we compare the theoretical models in a set of Markov Chain Monte Carlo (MCMC) analyses. We begin by exploring how much data from the mildly non-linear scales we can include without incurring in a significant bias on the measurement of the dark energy parameters. We do this for each perturbative model to select the best one -- the model which can explain the most data and extract the most information about the parameters of interest. Finally, we use that model to forecast constraints on the interaction parameter by stage IV surveys for the first time.

The paper is organised as follows. In Section~\ref{sec:model}, we describe the Dark Scattering model which we study in this work, as well as the three perturbative models for the non-linearities, whose performance we compare. In Section~\ref{sec:mcmc}, we present the details of our MCMC analysis and its set-up. Our results are shown in Sub-section~\ref{ssec:res}. Finally, our conclusions are presented in Section~\ref{sec:conc}.

\section{Modelling}\label{sec:model}

The cosmological model being considered is an interacting dark energy model in which dark energy exchanges momentum with dark matter without the accompanying energy exchange~\cite{Simpson:2010vh, Baldi:2014ica, Baldi:2016zom, Bose:2017jjx}. This is a phenomenological model inspired by the Thomson interaction between photons and charged particles, which is active in pre-recombination times. In that case, the energy exchange is suppressed after $e^+e^-$ annihilation because the typical energy carried by photons at this time is much smaller than the electron mass. Similarly, in the case under study it is the dark matter mass that is assumed to be much larger than the energy transferred in typical interactions with the dark energy during the late Universe. This gives rise to the following Euler equations for the velocity field of dark matter (c) and dark energy (DE),
\begin{align}
    \theta_{\rm c}'+\Hh\theta_{\rm c}+\n^2\phi=(1+w) \frac{\rho_{\rm DE}}{\rho_{\rm c}}an_{\rm c}\sigma_D(\theta_{\rm DE}-\theta_{\rm c})\,,\\
    \theta_{\rm DE}'-2\Hh\theta_{\rm DE}-\frac1{1+w}\n^2\delta_{\rm DE}+\n^2\phi=an_{\rm c}\sigma_D(\theta_{\rm c}-\theta_{\rm DE})\,,
\end{align}
in which a prime denotes a derivative with respect to conformal time, $\theta_i$ are the velocity divergences of species $i$, $\rho_i$ are their energy densities, $\delta_i$ are their density contrasts, $\Hh=a'/a$ is the conformal Hubble rate, $a$ is the scale factor, $\phi$ is the gravitational potential (in Newtonian gauge), $w$ is the dark energy equation of state parameter, $n_{\rm c}$ is the number density of dark matter and $\sigma_D$ is the cross-section of the interaction. Given that this model will be applied only at late times and no energy transfer occurs, no other equations are modified by the interaction. In addition to this, we employ the common approximation that dark energy fluctuations are negligible with respect to all others, given the assumed sound speed of $c_s=1$. Taking that into account, the resulting equations make it clear that the interaction introduces only an additional drag force to cosmic expansion and, converting time variables to scale factor, $a$, one finds
\begin{equation}
    a\p_a\Theta+\left(2+ (1+w)\xi \frac{\rho_{\rm DE}}{H}+\frac{a\p_aH}{H}\right)\Theta+\frac{\n^2\phi}{a^2H^2}=0\, ,\label{EqIDEmain}
\end{equation}
where the parameter $\xi\equiv\sigma_D/m_{\rm c}$ was introduced, $H=\dot{a}/a$ is the Hubble rate, with a dot representing a derivative with respect to cosmic time, and $\Theta\equiv\theta_{\rm c}/aH$.

This additional drag term is the only effect this model has at late times and is therefore a simple extension to $w$CDM, adding only one extra parameter, $\xi$.

\subsection{Evolution equations}

We assume the late-time evolution of the Universe is described by a perturbed flat Friedmann-Lema\^{i}tre-Robertson-Walker (FLRW) spacetime, whose metric, in Newtonian gauge, is given by the line element
\be
ds^2=-(1+2\phi) dt^2 + a^2(1-2 \psi)\delta_{ij} dx^idx^j\,,
\ee
where $\psi$ is the curvature perturbation and we have also neglected vector and tensor perturbations, which is a good approximation when predicting the power spectrum of density fluctuations. See Ref.~\cite{Malik:2008im} for a review of cosmological perturbation theory.

We assume the Universe to be composed only of dark matter and dark energy,\footnote{This approximation is likely to be slightly worse for this model than for $\Lambda$CDM, given that baryons are not expected to interact with dark energy, implying we are slightly overestimating the effect of the additional drag term. This is a small effect and only changes the effective value of the coupling. We leave a detailed analysis for future work.} with the latter having a constant equation of state. The background evolution is thus given by the Friedmann equation,
\be
H^2=H_0^2\left(\Omega_{c,0}a^{-3}+\Omega_{\rm{DE},0}a^{-3(1+w)}\right)\,,
\ee
where we have used the solution to the conservation of the background stress-energy tensors,
\be
\rho_{\rm{DE}}=\rho_{\rm{DE},0}\,a^{-3(1+w)}\,,\ \ \rho_{\rm{c}}=\rho_{\rm{c},0}\,a^{-3}\,.
\ee

The gravitational equations are obtained by applying the Newtonian limit to GR, resulting in 
\be
\n^2\phi=\frac32a^2H^2\Omega_{\rm c}\delta_{\rm c}\,,
\ee
where we have assumed negligible anisotropic stress, implying that $\psi=\phi$. 

Dropping the subscript $c$, the full non-linear fluid equations for dark matter in Fourier space are given by
\begin{align}
    &a\p_a\delta(\mathbf{k})+\Theta(\mathbf{k})=-\int{\frac{\rm{d}^3\mathbf{k_1}\rm{d}^3\mathbf{k_2}}{(2\pi)^3}\delta^{(3)}(\mathbf{k}-\mathbf{k_1}-\mathbf{k_2})}\alpha(\mathbf{k_1},\mathbf{k_2})\Theta(\mathbf{k_1})\delta(\mathbf{k_2})\,,\\
    &a\p_a\Theta(\mathbf{k})+\left(2+(1+w)\xi \frac{\rho_{\rm DE}}{H}+\frac{a\p_aH}{H}\right)\Theta(\mathbf{k})-\left(\frac{k}{aH}\right)^2\phi(\mathbf{k})=\nonumber\\
    &\quad\quad\quad\quad\quad\quad\quad-\frac12\int{\frac{\rm{d}^3\mathbf{k_1}\rm{d}^3\mathbf{k_2}}{(2\pi)^3}\delta^{(3)}(\mathbf{k}-\mathbf{k_1}-\mathbf{k_2})}\beta(\mathbf{k_1},\mathbf{k_2})\Theta(\mathbf{k_1})\Theta(\mathbf{k_2})\,,
\end{align}
where the couplings $\alpha$ and $\beta$ are given by
\be
\label{eq:modecoupling}
\alpha(\mathbf{k_1},\mathbf{k_2})=1+\frac{\mathbf{k_1}\cdot\mathbf{k_2}}{|\mathbf{k_1}|^2}\,,\ \ 
\beta(\mathbf{k_1},\mathbf{k_2})=\frac{\mathbf{k_1}\cdot\mathbf{k_2}|\mathbf{k_1}+\mathbf{k_2}|^2}{|\mathbf{k_1}|^2|\mathbf{k_2}|^2}\,.
\ee

These equations can be solved perturbatively in terms of the kernels $\bar F_n$, $\bar G_n$ defined by
\begin{align}
    \delta(\mathbf{k},a)=\sum_{n=1}^{\infty}\int\frac{\rm{d}^3\mathbf{k_1}\dots\rm{d}^3\mathbf{k_n}}{(2\pi)^{3(n-1)}}\delta^{(3)}(\mathbf{k}-\mathbf{k_{1n}})\bar F_n(\mathbf{k_1},\dots,\mathbf{k_n},a)\delta_0(\mathbf{k_1})\dots\delta_0(\mathbf{k_n})\,,\\
    \Theta(\mathbf{k},a)=\sum_{n=1}^{\infty}\int\frac{\rm{d}^3\mathbf{k_1}\dots\rm{d}^3\mathbf{k_n}}{(2\pi)^{3(n-1)}}\delta^{(3)}(\mathbf{k}-\mathbf{k_{1n}})\bar G_n(\mathbf{k_1},\dots,\mathbf{k_n},a)\delta_0(\mathbf{k_1})\dots\delta_0(\mathbf{k_n})\,,
\end{align}
where $\delta_0$ is the initial density contrast for the growing mode solution defined deep in matter domination, $\mathbf{k_{1n}}=\sum_{i=1}^n \mathbf{k_i}$, and the kernels $\bar F_n$, $\bar G_n$ can be constructed with recursive relations from
the fundamental mode coupling functions~\eqref{eq:modecoupling}~\cite{Jain:1994}. The first order density kernel $\bar F_1(a)$ is the growth factor, which we denote by $D(a)$. We also define the growth rate $f\equiv{\rm d}\ln{D}/{\rm d}\ln{a}=-\bar G_1/\bar F_1$, parameterising the linear growth of velocities. The higher order kernels are computed order by order by using the Einstein-de Sitter (EdS) approximation, i.e. they are calculated for a matter dominated Universe and then their time evolution is corrected by substituting the EdS growth factors by the linear growth factors computed in the full $w$CDM interacting model. In particular, this implies, for $n\geq2$,
\begin{align}
\bar F_n(\mathbf{k_1},\dots,\mathbf{k_n},a)&=D^n F_n(\mathbf{k_1},\dots,\mathbf{k_n})\,,\\
\bar G_n(\mathbf{k_1},\dots,\mathbf{k_n},a)&=- f D^n  G_n(\mathbf{k_1},\dots,\mathbf{k_n})\,,
\end{align}
with $F_n$ and $G_n$ being the time-independent EdS kernels~\cite{Bernardeau:2001qr}. This is a good approximation given the scale-independent nature of the linear growth in this model. The approximation has also been explicitly tested against the full solutions, and is found to be in sub-percent agreement at the scales and redshifts of interest in this paper.

\subsection{Dark matter power spectrum modelling}

We study different models for the mildly non-linear scales that are based on perturbation theory, which we construct here step-by-step. The first ingredient is the dark matter power spectrum, whose first building blocks for all models under study are the standard perturbation theory (SPT) 1-loop power spectra, given by
\be
P^{1-{\rm loop}}_{ij}(k;a)  = F_{ij}\Big[ D^2P_L(k) + D^4P^{22}_{ij}(k) + D^4P^{13}_{ij}(k)\Big],
\label{eq:1loop}
\ee
where $i,j \in \{\delta,\Theta\}$ and $F_{\delta \delta} = 1, F_{\delta \Theta} = -f$ and $F_{\Theta \Theta}=f^2$. $P_L$ denotes the power spectrum of the initial density contrast and the remaining components are loop integrals over the EdS kernels, which are given by
\begin{align}
P_{\delta \delta}^{22}(k)
&= \frac{2}{(2\pi)^{3}}\int \de^3 q F_2 (\bfk - \bfq, \bfq)^2 
P_L(|\bfk - \bfq|) P_L(q), \\
P_{\delta \Theta}^{22}(k) 
&= \frac{2}{(2\pi)^{3}} \int \de^3 q F_2 (\bfk - \bfq, \bfq) G_2 (\bfk - \bfq, \bfq)
P_L(|\bfk - \bfq|) P_L(q), \\
P_{\Theta \Theta}^{22}(k) 
& = \frac{2}{(2\pi)^{3}} \int \de^3 q G_2 (\bfk - \bfq, \bfq)^2
P_L(|\bfk - \bfq|) P_L(q),
\end{align}
and 
\begin{align}
P_{\delta \delta}^{13}(k) 
&= \frac{6}{(2\pi)^{3}} \int \de^3 q F_3(\bfk, \bfq, -\bfq) 
P_L(q) P_L(k), \\
P_{\delta \Theta}^{13}(k) 
&= \frac{3}{(2\pi)^{3}}  \int \de^3 q G_3 (\bfk, \bfq, -\bfq)
P_L(q)  P_L(k) + 3  \int \de^3 q F_3 (\bfk, \bfq, -\bfq) 
P_L(q) P_L(k), \\
P_{\Theta \Theta}^{13}(k) 
& = \frac{6}{(2\pi)^{3}} \int \de^3 q G_3 (\bfk, \bfq, -\bfq) 
P_L(|\bfk - \bfq|) P_L(q).
\end{align}

The validity of the 1-loop SPT power spectrum can be further extended in the non-linear regime by adopting the EFTofLSS prescription.
This formalism systematically accounts for the impact of non-linearities on mildly non-linear scales via the introduction of effective stresses in the equations of motion~\cite{Baumann:2010tm,Carrasco:2012cv}. This amounts to renormalising the fields under consideration, such as $\delta$ and $\Theta$, which results in the addition of counter-terms to the 1-loop power spectrum:
\be
P_{ij}^{\rm EFT}(k; a)=P^{1-{\rm loop}}_{ij}(k;a)-2 D^2 c_{ij}(a) k^2 P_L(k)\,,
\ee
with 
\begin{align}
c_{\delta\delta}(a)&=c_{2|\delta}(a)\,,\nonumber\\ 
c_{\delta\Theta}(a)&=-\frac{f}{2} (c_{2|\delta}(a)+c_{2|\Theta}(a))\,,\label{c2dt}\\ 
c_{\Theta\Theta}(a)&=f^2 c_{2|\Theta}(a)\,,\nonumber
\end{align}
where $c_{2|\delta}$ and $c_{2|\Theta}$ are the counter-terms used to renormalise the fields $\delta$ and $\Theta$ up to $O((k/k_{\rm NL})^2)$, $k_{\rm NL}$ being the rough scale at which non-linearities become important.

\subsection{Bias model}

We consider a general bias expansion to relate the galaxy density field to the underlying dark matter~\cite{McDonald:2009dh,Assassi_2014,Desjacques:2016bnm}: 
\be
\delta_g = b_1 \delta + \frac{b_2}{2} \delta^2 + b_{\gamma_2} \mathcal{G}_2 + b_{\Gamma_3} \Gamma_3 + b_{\nabla^2 \delta} \nabla^2 \delta +\epsilon,
\ee
where $b_1$ and $b_2$ are the linear and quadratic bias, $\mathcal{G}_2$ and $\Gamma_3$ are non-local operators, $\nabla^2 \delta$ is a higher derivative operator and $\epsilon$ is the stochastic contribution encoding deviations from Poisson shot-noise. The latter contribution is uncorrelated with the other fields and we approximate it to have a scale-independent power spectrum $P_{\epsilon\epsilon}\equiv N$, which is one of our bias parameters. In this work we reduce the bias parameter space by setting $b_{\Gamma_3}$ to zero, similarly to what was recently done in the re-analysis of BOSS data~\cite{Ivanov:2019pdj}. 
The higher derivative contribution to the power spectrum given by $b_{\nabla^2 \delta}$ should be important only for very biased tracers~\cite{Fujita:2020}, therefore we neglect it in our analysis. We also note that, for the EFTofLSS models, its contribution is perfectly degenerate with the EFTofLSS counterterm. Those models can therefore take into account also the $b_{\nabla^2 \delta}$ contribution via a re-definition of the $c_{2|\delta}$ parameter. This results in the following power spectrum:
\begin{align}
    P_{gg}(k) =& \, b_1^2 P^{1-{\rm loop}}_{\delta \delta} (k) + D^4\Big[ 2 b_1 \left(b_2-\frac43 b_{ \gamma_2}\right) P_{b2,\delta}(k) +4 b_1 b_{ \gamma_2} P_{bs2,\delta}(k) + \frac{32}{45} b_1 b_{\gamma_2}\sigma_3^2(k)P_L(k) \nonumber \\ & + \left(b_2-\frac43 b_{ \gamma_2}\right)^2 P_{b22}(k) + 4 \left(b_2-\frac43 b_{ \gamma_2}\right) b_{\gamma_2} P_{b2s2}(k)  + 4 b_{\gamma_2}^2 P_{bs22} (k)\Big] + N, \label {pb1} \\
P_{g \Theta}(k) =&\,  b_1 P^{1-{\rm loop}}_{\delta \Theta} (k) - f D^4\Big[ \left(b_2-\frac43 b_{ \gamma_2}\right)  P_{b2,\Theta}(k) +2 b_{\gamma_2} P_{bs2,\Theta}(k) \nonumber\\
&\quad\quad\quad\quad\quad\quad\quad\quad\quad\quad\quad\quad\quad\quad\quad\quad\quad\quad\quad\quad\quad+ \frac{16}{45}b_{\gamma_2}\sigma_3^2(k)P_L(k) \Big], \label{pb2}
\end{align}
in which the terms $P_{b...}$ above are given by Eqs. (14) to (22) of \cite{Bose:2019psj}.
The parameter space can be further reduced by adopting the local-Lagrangian relation~\cite{Chan:2012jj,Baldauf:2012hs,Sheth:2012fc,Saito:2014qha}, given by
\be
b_{\gamma_2}=-\frac27(b_1-1)\,,\label{lLb}
\ee
allowing us to write the real-space galaxy power spectrum as a function of only three time-dependent bias parameters. We assume there is no velocity bias. 

\subsection{Redshift-space modelling}

To model the redshift-space power spectrum of biased tracers, we make use of three different models: the Taruya, Nishimici, Saito (TNS) model~\cite{Taruya:2010mx}, and two EFT-based models: EFT-I, presented in \cite{Bose:2019psj} and based on the model of \cite{delaBella:2017qjy}, and EFT-II, described in \cite{Chudaykin:2020} and based on the work of \cite{Perko:2016puo}.
The TNS model adopts the local Lagrangian relation, Eq.~\eqref{lLb}, as does the EFT-I model. The two EFTofLSS models share the same loop corrections, but adopt different counterterms and prescriptions for the Fingers-of-God, which we describe in more detail below.\\\\

\noindent The TNS power spectrum can be written as
\begin{align}
 P_{\rm TNS}(k,\mu) =& D_{\rm FoG}(\mu^2 k^2 \sigma_v^2)\Big[ P_{gg} (k) - 2 \mu^2 P_{g \Theta}(k) +  \mu^4 P_{\Theta \Theta}^{\rm 1-loop} (k) \nonumber \\ & + b_1 A(k,\mu) + b_1^2 B(k,\mu) + C(k,\mu)  \Big], 
\label{redshiftpsTNS}
\end{align} 
where $\mu=\hat\bfk\cdot\hat \bfn$ is the cosine of the angle between the line of sight direction, $\hat\bfn$, and the wave-vector, $\bfk$. The terms $A$, $B$, $C$ are redshift-space corrections from SPT, which are given by Eqs. (11) to (13) of \cite{Bose:2019psj}, with further details given in \cite{Taruya:2010mx}.

The key feature of this model is the overall factor $D_{\rm FoG}$, which encodes a phenomenological description of the Fingers-of-God (FoG) damping. This is a function of the velocity dispersion, $\sigma_v$, which is taken to be a free parameter. We assume this function takes a Lorentzian form so that
\be
D_{\rm FoG}(k^2\mu^2 \sigma_v^2) = \frac{1}{1 + (k^2\mu^2 \sigma_v^2)/2} \, ,\label{DFoG}
\ee
which has been shown to be accurate at describing this effect~\cite{Bose:2019psj}. Given this additional parameter for the FoG damping, the full set of 4 nuisance parameters for this model is $\{b_1,b_2,N,\sigma_v\}$.\\\\

The second model we study is based on the EFTofLSS and we name it EFT-I. The first contribution is the full SPT power spectrum in redshift space, given by
\begin{align}
 P_{\rm SPT}(k,\mu) =&  P_{gg} (k) - 2 \mu^2 P_{g \Theta}(k) +  \mu^4 P_{\Theta \Theta}^{\rm 1-loop} (k) \nonumber \\ & + b_1 A(k,\mu) + b_1^2 B(k,\mu) + C(k,\mu) - D^4 f^2 k^2\mu^2 \tilde\sigma_v^2 (b_1+f\mu^2)^2 P_L(k)
 \, , 
\label{redshiftpsSPT}
\end{align}
with the SPT estimate for the velocity dispersion, $\tilde\sigma_v$, given by
\begin{equation}
\tilde{\sigma}^2_{v} = \frac{1}{6\pi^2} \int {\rm d}q P_L(q) \ .
\end{equation}
As was the case for the TNS model, this parameter and its corresponding term represent the FoG damping, but in this case, it is fixed to the SPT prediction instead of being a model parameter. Indeed, expanding Eq.~\eqref{DFoG} up to $O(k^2)$ reveals a very similar term to the last term in Eq.~\eqref{redshiftpsSPT}, should one relate the two velocity dispersions via  $\sigma_v^2=2 D^2 f^2 \tilde{\sigma}^2_{v}$.

We apply a re-summation procedure to model the non-linear damping of the BAO feature in the power spectrum, following \cite{Vlah:2015zda,delaBella:2017qjy}. We perform a decomposition of the linear power spectrum between the smooth broadband ($P^{\rm nw}$) and wiggle ($P^{\rm w}$) components as given by Eq.~(2.47) of \cite{delaBella:2017qjy}, using the Eisenstein-Hu fitting function \cite{Eisenstein:1997ik}, and introduce a damping factor
\be
\bar{\Sigma}^2= \frac12 [1+f(f+2)\mu^2] \frac{D^2}{\pi^2} \frac{1}{q_{\rm max}^3 - q_{\rm min}^3} \int_{q_{\rm min}}^{q_{\rm max}} \de q q^2 \int_0^{\infty}\de k P_L^{\rm nw} (k)\left[1-j_0(qk)\right]\,,
\ee
where $j_0$ is the spherical Bessel function,  $q_{\rm max}=(300\ {\rm Mpc}/h)$ and $q_{\rm min}=(10\ {\rm Mpc}/h)$, which roughly delimit the region over which $P^{\rm w}$ is non-zero. 

The re-summed next-to-leading order power spectrum is then given by
\be
P_{\rm SPT}^{\rm re-sum}=P_{\rm SPT}^{\rm nw} + e^{-k^2\bar{\Sigma}^2}  P_{\rm SPT}^{\rm w} + D^2 e^{-k^2 \bar{\Sigma}^2}k^2 \bar{\Sigma}^2(b_1+f\mu^2)^2 P_L^{\rm w}(k)\,. \label{eq:resum}
\ee

Finally, we add the EFTofLSS counter-terms encoding the effects of non-linearities, which gives the final power spectrum for this model as
\be
P_{\rm EFT}=P_{\rm SPT}^{\rm re-sum} - 2D^2P_L(k)k^2\Big[c^2_{s,0} + {c}^2_{s,2} \mu^2 + {c}^2_{s,4} \mu^4 + \mu^6 \left(f^3 {c}^2_{s,0} - f^2 {c}^2_{s,2} + f {c}^2_{s,4}\right)\Big]\,,
\label{redshiftpsEFT}
\ee
where the parameters $c_{s,i}^2$ are coefficients of an expansion in powers of $\mu^2$ and are combinations of the EFTofLSS parameters of Eq.~\eqref{c2dt} with further parameters used to renormalize composite operators required in redshift space. This set of 3 EFTofLSS parameters represents a complete set of possible contributions at 1-loop order and up to order $O((k/k_{NL})^2)$. We do not include stochastic counter-terms, as these are either degenerate with the stochastic term included in the bias model or are effectively $k^4$ contributions~\cite{delaBella:2017qjy,Schmittfull:2018yuk}, which we choose not to include in this model. Given the above, this model has an additional 2 parameters than the TNS model, resulting in a set of 6 nuisance parameters: $\{b_1,b_2,N,c_{s,0}^2,c_{s,2}^2,c_{s,4}^2\}$.\\\\

The third and final model we consider is called here EFT-II and is the one described in \cite{Ivanov:2019pdj,Chudaykin:2020}, that assumes the Standard Perturbation Theory one-loop expression (see, e.g.~\cite{Bernardeau:2001qr}) with the addition of EFTofLSS counterterms. The loop corrections are computed from the redshift-space kernels $Z_1(\bfk)$, $Z_2(\bfk_1, \bfk_2)$, $Z_3(\bfk_1, \bfk_2, \bfk_3)$ (see~\cite{Scoccimarro:1999} and Eq.~(2.14) of~\cite{Ivanov:2019pdj} for a derivation with the same bias basis we adopt). From these kernels, the redshift space galaxy power spectrum at one-loop can be written as
\begin{align}
\label{eq:eftii}
P_{gg}(\bfk) &= Z_1^2(\bfk) P_L(k) + 2 \int \de^3 \bfq \, \left[ Z_2(\bfq, \bfk - \bfq) \right]^2P_L(q) P_L(\vert \bfk - \bfq \vert) \nonumber \\
&+ 6 Z_1(\bfk)P_L(k) \int \de^3 \bfq \, Z_3(\bfk, \bfq, -\bfq)P_L(q) + P_{\rm ctr}(\bfk) + N\,,
\end{align}
where the loop integrals are computed over the IR-resummed linear power spectrum, defined by the linear version of Eq.~\eqref{eq:resum}. The redshift-space loop corrections lead to 28 independent integrals, which we compute using the {\sc Fast-PT} algorithm~\cite{McEwen:2016,Fang:2017}.
The EFTofLSS counterterms for this model are
\be
P_{ctr}(k,\mu) = - 2 \tilde{c}_0 k^2 P_L(k) - 2 \tilde{c}_2 k^2 f \mu^2 P_L(k) - 2 \tilde{c}_4 k^2 f^2 \mu^4 P_L(k) +P_{ctr, \nabla^4 \delta}(k,\mu),
\ee
with an additional counterterm proportional to  $\mu^4 k^4 P_L(k)$ to include higher-order contributions and model to some extent the FoG:
\be
P_{ctr, \nabla^4 \delta}(k,\mu) = c_{\nabla^4 \delta} k^4 f^4 \mu^4 (b_1 + f \mu^2)^2 P_L(k) .
\ee
This model has therefore two more parameters than the EFT-I model, resulting in 8 nuisance parameters: $\{b_1, b_2, b_{\gamma_2}, N, \tilde{c}_0, \tilde{c}_2, \tilde{c}_4, c_{\nabla^4 \delta}\}$.\\\\

\noindent To summarize, the main differences between the three models considered here are:
\begin{itemize}
    \item \textbf{Perturbation theory prescription}: The TNS model is based on standard perturbation theory, while EFT-I and EFT-II are based on the EFTofLSS.
    \item \textbf{Bias model}: TNS and EFT-I assume the local Lagragian relation, Eq.~\eqref{lLb}, while the EFT-II model does not. The simplest models therefore include 3 free bias parameters: linear bias, $b_1$, second-order bias, $b_2$, and the amplitude of the stochastic noise spectrum, $N$; while the EFT-II model includes additionally the tidal bias, $b_{\gamma_2}$.
    \item \textbf{Fingers-of-God damping}: TNS uses a phenomenological factor multiplying the entire redshift-space power spectrum, EFT-I uses the 1-loop prediction from perturbation theory and EFT-II uses an additional $O(k^4)$ counterterm for this purpose. The $O(k^2)$ counterterms in both EFT models also somewhat contribute to the description of this effect.
    \item \textbf{Counterterms}: TNS has no counterterms, both EFT models have 3 $O(k^2)$ counterterms, with EFT-I including contributions up to $O(\mu^6)$, while EFT-II only going up to $O(\mu^4)$, but including an additional $O(k^4)$ counterterm.
\end{itemize}

All three models have been previously used in the literature in similar versions to the ones described here. The TNS model has been used in the analysis of BOSS data~\cite{Beutler:2013yhm,Beutler_2016}, albeit using renormalised perturbation theory instead of SPT, without including the $C$ term and using a Gaussian prescription for the FoG damping, instead of the improved Lorentzian one used here~\cite{Bose:2019psj}. The EFT-I model has been used in \cite{Markovic:2019sva,Bose:2019psj,Bose:2019ywu}, albeit using a partial re-summation prescription. Both of these models were analysed in \cite{delaBella:2018fdb}, with the TNS model being similar to their \verb|SPT+Coevo| model and the EFT-I model being similar to their \verb|EFT+Coevo| model. The EFT-II model has also been used to analyse the BOSS data in \cite{Ivanov:2019pdj}, as well as in \cite{DAmico:2019fhj}, but using a more comprehensive bias model, including higher order stochastic terms. Those two versions of this model have been compared in \cite{Nishimichi:2020tvu} in a blinded challenge, which showed good agreement and the ability of this model to accurately describe high precision simulations.

\section{MCMC analysis}\label{sec:mcmc}

We now present our MCMC analysis of the interacting dark energy model described in Section~\ref{sec:model}. We compute forecasts for the the dark energy equation of state parameter, $w$, as well as for the DM-DE interaction parameter, defined by
\be
A\equiv \xi (1+w)\,.
\ee
We choose this combination of parameters instead of $\xi$ as it more directly encodes the strength of the effect of the interaction on the observables under consideration, as can be inferred from Eq.~\eqref{EqIDEmain}.\footnote{A similar variable has been used in other interacting dark energy models in Ref.~\cite{Yang:2018euj}, for similar reasons.} Indeed, it is possible to constrain this parameter even when $w\approx-1$, whereas $\xi$ can take very large values in that scenario. Note that the two parameters of interest, $w$ and $A$, affect the evolution of the Universe in distinct ways, since $w$ changes both the background cosmology and the perturbations, whereas $A$ can only affect the latter. It is this fact that may allow this model to fit data very well and alleviate apparent cosmological tensions. 

We base our forecasts on synthetic observations of the multipoles of the galaxy-galaxy power spectrum in redshift space, defined as
\be
P_\ell(k)=\frac{2\ell+1}{2}\int_{-1}^1{\rm d} \mu P(k,\mu)\mathcal{P}_\ell(\mu)\,,
\ee
where $\mathcal{P}_\ell(\mu)$ are the Legendre polynomials of order $\ell$ and $P(k,\mu)$ is given by the theoretical prediction from one of the models of Eq.~\eqref{redshiftpsTNS}, Eq.~\eqref{redshiftpsEFT} or Eq.~\eqref{eq:eftii}. In this analysis we use the monopole ($\ell=0$), quadrupole ($\ell=2$) and hexadecapole ($\ell=4$). 

Our mock data is composed of four Parallel COmoving Lagrangian Acceleration (PICOLA) simulations~\cite{Howlett:2015hfa,Winther:2017jof} of a $\Lambda$CDM cosmology with box length 1024 Mpc/$h$ with 1024$^3$ dark matter particles evolved from initial redshift $z_{\rm ini}=49$. We use two snapshots at redshifts $z=0.5$ and $z=1$. Given our focus on biased tracers, we use halo catalogs, built using the friends-of-friends algorithm with a linking length of 0.2-times the mean particle separation. We employ a mass cut of $M_{\rm min}=4\times10^{12}~M_\odot$, chosen to fix the number density of halos to $n_{\rm h}=10^{-3}~h^3/{\rm Mpc}^3$. The combined volume of these simulations as well as the chosen number density make this an appropriate set of mock data to approximate the properties of stage IV surveys, such as Euclid and DESI~\cite{Blanchard:2019oqi, Aghamousa:2016zmz},
for observations near the considered redshifts. Measurements of the power spectrum multipoles from these simulations are made using the distant observer approximation, and then averaged over three lines of sight as well as over the four simulations. These are the same simulations used in the analysis of \cite{Markovic:2019sva,Bose:2019psj}.

Our $\Lambda$CDM simulations use the WMAP9 cosmology~\cite{Hinshaw:2012aka}, with cosmological parameters $\Omega_m=0.281$, $\Omega_b=0.046$, $h=0.697$, $n_s=0.971$ and $\sigma_8(z=0)=0.844$.\footnote{The reason for choosing this cosmology is due to the prompt availability of these simulations. We do not expect that using a fiducial cosmology from a more recent survey would alter any of our results, given that the changes in parameter values are relatively minor.} Using these naturally assumes fiducial values for the parameters of interest of $w=-1$ and $A=0$ and allows us to forecast the best constraints possible on these parameters should the true cosmology be $\Lambda$CDM. 

The main aim of this work is to produce accurate forecasts for the dark energy parameters. This can only be done after a careful analysis of the modelling of the mildly non-linear scales. Fortunately, our MCMC analysis allows us to do both. We will first test our modelling via the TNS and EFTofLSS models presented above by measuring the bias in the measurement of $w$ or $A$ with respect to the fiducial values of the simulations, since this is a clear diagnostic of the accuracy of these models. This allows us to determine the maximum wave-number, $k_{\rm max}$, at which the models considered give sufficiently accurate descriptions of the power spectrum multipoles. After this procedure, we are able to choose the most complete model and the appropriate scales to include in order to obtain the best unbiased forecasts, which will be our final result.

\subsection{MCMC set-up}

We now describe the set-up of our MCMC analysis. Our aim is to analyse the mock data at redshifts $z=0.5$ and $z=1$ with the purpose of measuring the parameters of the interacting dark energy model, $w$ and $A$, leaving the nuisance parameters free to vary, but fixing the other cosmological parameters to the simulation's fiducial value. 

We approximate the likelihood to be Gaussian and given by
\be
-\ln \mathcal{L} \propto \chi^2 =  \sum_{\ell,\ell'=0,2,4}~ \sum_{k=k_{\rm min}}^{k_{\rm max}^{\ell,\ell'}}\left[P_{\ell}^{\rm dat}(k)-P_{\ell}^{\rm mod}(k)\right] \mbox{Cov}^{-1}_{\ell,\ell'}(k)\left[P^{\rm dat}_{\ell'}(k)-P_{\ell'}^{\rm mod}(k)\right],
\label{chi2}
\ee
where the minimum wave-number is $k_{\rm min}=2\pi / {\rm L_{box}} = 0.006~h/{\rm Mpc}$ and the maximum wave-number considered is denoted as $k_{\rm max}^{\ell,\ell'}$ to make it explicit that it is multipole-dependent. Here we take the same $k_{\rm max}$ for monopole and quadrupole, but assume different values for the hexadecapole, as is done in real data analyses, such as those of the BOSS data (see, e.g., \cite{Beutler:2013yhm,Beutler_2016,DAmico:2019fhj,Ivanov:2019pdj}). We explore how the variation of both wave-numbers introduces a bias in the best fit dark energy parameters with respect to the fiducial values. The covariance matrix, $\mbox{Cov}_{\ell,\ell'}$, is calculated assuming the fluctuations to be Gaussian and based on the Kaiser power spectrum calculated in the fiducial cosmology. In addition, we assume a number density of $n=10^{-3}~h^3/{\rm Mpc}^3$ and a survey volume of $4~{\rm Gpc}^3/h^3$ at both redshifts in the calculation of the covariance matrix, matching the properties of our simulated data as well as those of stage IV surveys.

Regarding priors, we place a positivity prior on the nuisance parameter $\sigma_v$ of the TNS model, as well as a prior on the dark energy parameters such that $\xi=A/(1+w)\geq0$. This is because the parameter $\xi$ is the ratio of a cross section and a mass, both of which have to be positive.

For each $k_{\rm max}$, we perform a least-squares fit to find suitable parameter values to initialize the MCMC chains. This is repeated three times with random variations of the initial conditions and the point with the smallest $\chi^2$ is chosen. MCMC chains are then run using Goodman \& Weare’s affine invariant sampler implemented in \verb|emcee|~\cite{Foreman_Mackey_2013}. We follow this procedure for a range of $k_{\rm max}^{\ell=0,2}\in [0.1,0.3]~h/{\rm Mpc}$ and $k_{\rm max}^{\ell=4}\in [0,k_{\rm max}^{\ell=0,2}]$\footnote{We only explore the inclusion of scales for the hexadecapole that are larger than those included in the monopole and quadrupole. This is because we expect the hexadecapole to be less well described by our modelling than the lower order monopoles and therefore including a broader range of scales for it would only bias results further. As we will see below, this is confirmed by our results.} to find the point at which the bias on the dark energy parameters exceeds their standard deviation. We follow a similar procedure to Refs.~\cite{Osato_2019,Eggemeier:2020umu,Pezzotta:2021vfn} and evaluate this via the so-called Figure-of-Bias (FoB), defined as
\be
{\rm FoB}(\theta)=\sqrt{(\bar\theta-\theta_{\rm fid})^T S^{-1}(\bar\theta-\theta_{\rm fid})}\,,
\ee
where $\theta=[w,A]$ is a vector with the dark energy parameters, $\bar\theta$ is the mean of their posterior, $\theta_{\rm fid}=[-1,0]$ is the fiducial parameter vector and $S$ is the parameter covariance matrix, calculated directly from the sampled posterior. This measure of bias is constructed from a re-scaling of the 2D contours so that they are more symmetric and have a radius of order unity. The FoB then measures how far from the centre of the contours the fiducial values are, in units of their standard deviation. In particular, for an exactly Gaussian distribution, this re-scaling is equivalent to the usual standardization and so the contours appear as an exact circle, and the FoB would evaluate to 1.52 and 2.49 at the limits of the 68\% and 95\% confidence regions, respectively, since these are the values obtained for a standardized 2D Gaussian. While our posterior is not Gaussian, we will use these values to estimate the confidence intervals and we set the threshold of acceptance of our modelling at the 68\% confidence level. 

In addition to this, we test the goodness of fit and quote the value of $\chi^2/N_{\rm dof}$ for the best-fit point as a function of $k_{\rm max}$, where the number of degrees of freedom, $N_{\rm dof}$, is given by
\be
N_{\rm dof}=N_k-N_{\rm par}\,,
\ee
with $N_k$ the total number of $k$-bins, and $N_{\rm par}$ the effective number of free parameters in the model. For the TNS model we always have $N_{\rm par}=6$, whereas for the EFTofLSS models, one of the counterterms cannot be constrained unless the hexadecapole is included. Therefore, for the analysis considering only the monopole and quadrupole, we have $N_{\rm par}=7$ for the EFT-I model and $N_{\rm par}=9$ for the EFT-II model, whereas the analysis including all multipoles has, respectively, $N_{\rm par}=8$ and $N_{\rm par}=10$. We use the expected standard deviation for the normalized $\chi^2$ distribution (with mean equal to 1) given by $\sigma=\sqrt{2/N_{\rm dof}}$ to classify the goodness of fit and set the threshold for acceptance at the 1$\sigma$ level.

Finally, we aim to forecast the best constraints possible on the parameters of interest. To quantify this, we use the figure-of-merit (FoM), given by
\be
{\rm FoM}(\theta)=\frac{1}{\sqrt{\det S}}\,.
\ee
As shown in Figs.~\ref{All_perf_z05} and \ref{All_perf_z1}, we compute this also as a function of $k_{\rm max}$ in order to establish the threshold after which no new information is gained from additional data.

\subsection{Results}\label{ssec:res}

\begin{figure}[h]
    \centering
		\includegraphics[height=0.3\textheight]{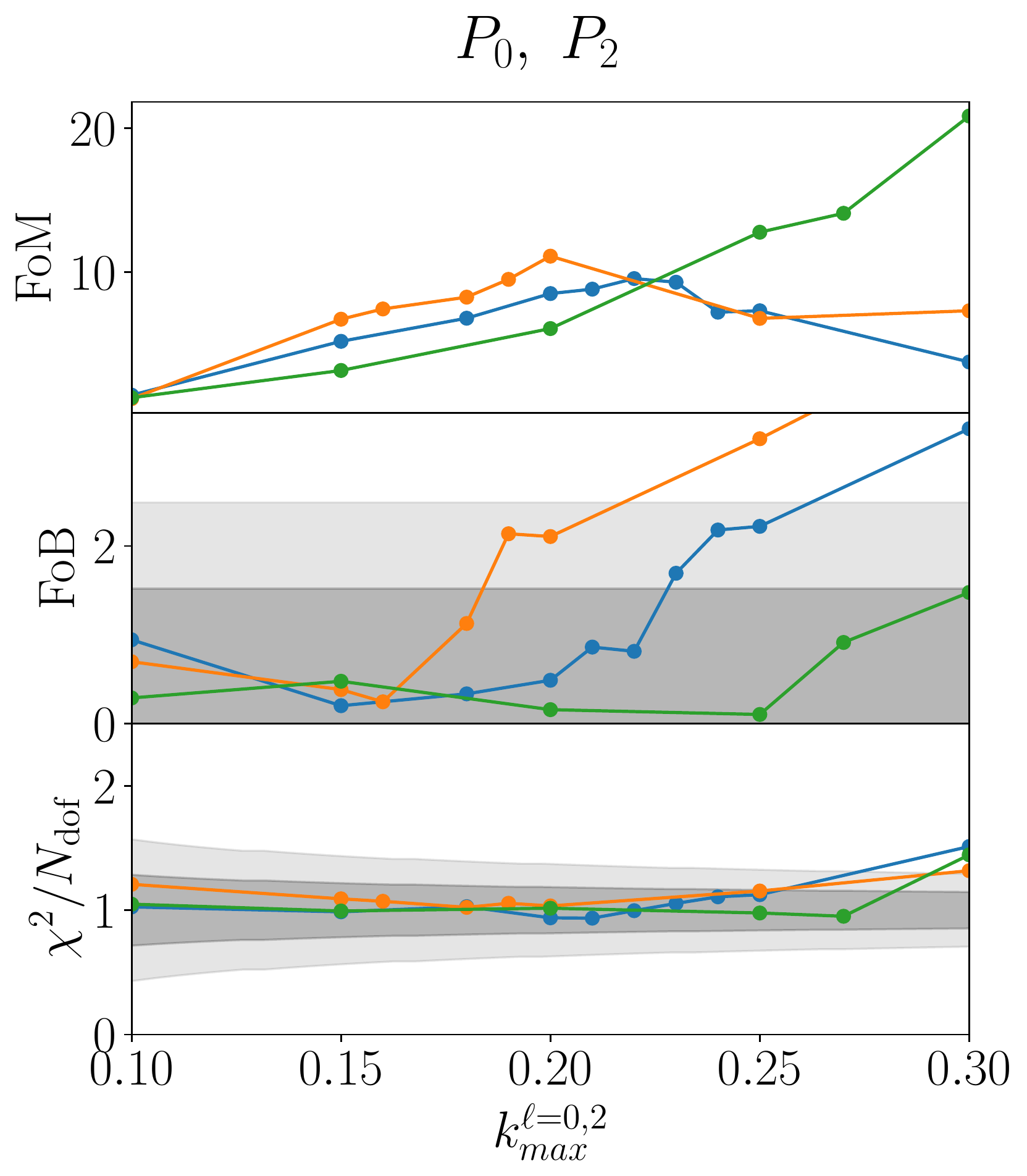}
		\includegraphics[height=0.3\textheight]{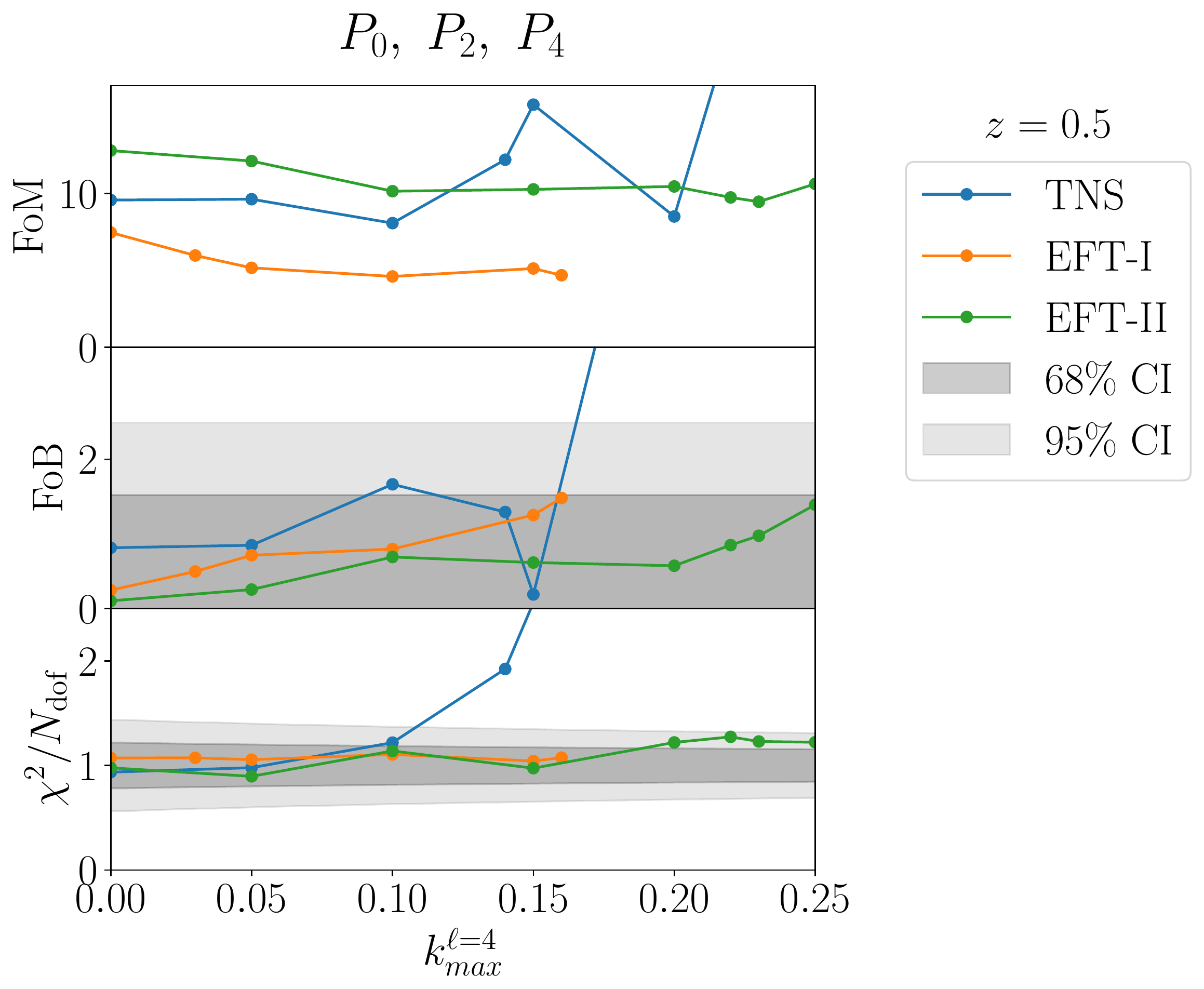}
    \caption{Performance measures for all three models for $z=0.5$. Left -- neglecting the hexadecapole, as a function of the maximum wave-number of the monopole and quadrupole. Right -- including the hexadecapole, as a function of its maximum wave-number, with the maximum wave-numbers for monopole and quadrupole being $k_{\rm max}=0.22~h/{\rm Mpc}$ for the TNS model, $k_{\rm max}=0.16~h/{\rm Mpc}$ for EFT-I and $k_{\rm max}=0.25~h/{\rm Mpc}$ for EFT-II. Confidence intervals shown for FoB are calculated from a two-dimensional Gaussian distribution and occur at ${\rm FoB}=1.52$ and $2.49$, for $68\%$ and $95\%$, respectively. Confidence intervals shown for the reduced-$\chi^2$ are estimated from the 1$\sigma$ and 2$\sigma$ distances away from the mean, with $\sigma=\sqrt{2/N_{\rm dof}}$ resulting from the $\chi^2$ distribution. As $N_{\rm dof}$ varies with each model, we chose to plot the broadest errors, corresponding to the EFT-II model for $P_0,~P_2$ and to the EFT-I model for $P_0,~P_2,~P_4$.}
    \label{All_perf_z05}
\end{figure}

We show our results for the performance measures of the three models considered for $z=0.5$ in Fig.~\ref{All_perf_z05}. From these results, we can see that the different models have varying ranges of validity, with EFT-I working well at small values of $k_{\rm max}$, while TNS is the better model at intermediate values and EFT-II having the largest reach, as well as figure-of-merit overall. When we include only the monopole and quadrupole, the maximum values of $k_{\rm max}^{\ell=0,2}$ for which both the bias and the reduced chi-squared are within their 68\% confidence intervals are given in Table~\ref{Mostkmax}. 
\begin{table}[h]
\centering
 \begin{tabular}{||c c c||} 
 \hline 
 Model &\ \ &$k_{\rm max}^{\ell=0,2}$ $(h/{\rm Mpc})$\\
 \hline\hline
 TNS &\ \ & 0.22 \\  
 \hline
 EFT-I &\ \ & 0.18 \\ 
 \hline
 EFT-II &\ \ & 0.27 \\ 
 \hline
\end{tabular}
\caption{Maximum values of $k_{\rm max}^{\ell=0,2}$ for which both the bias and the reduced chi-squared are within their 68\% confidence intervals, for each model for the analysis at $z=0.5$.}
\label{Mostkmax}
\end{table}
While it is the model with the largest biases, the EFT-I model has a larger FoM than the TNS model for smaller $k_{\rm max}^{\ell=0,2}$. When selecting only the points for which the bias is within the 68\% confidence interval, the results are similar between the two models, reaching a FoM of 8-10. The more complete EFT-II model can reach double that value, but at the cost of degrading the quality of the fit. With a good fit, this model can reach ${\rm FoM}=15$.

\begin{figure}[h]
    \centering
		\includegraphics[height=0.3\textheight]{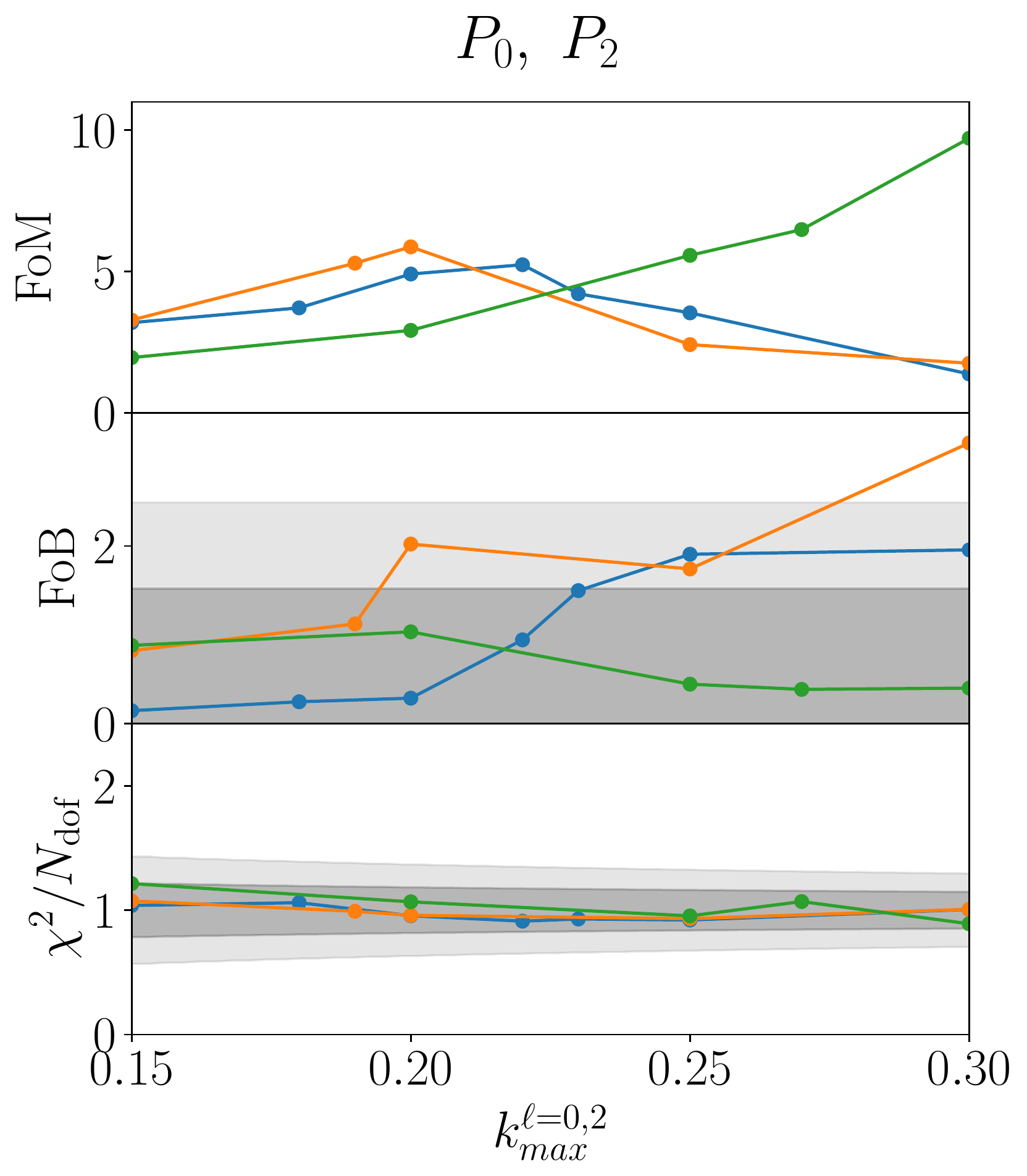}
		\includegraphics[height=0.3\textheight]{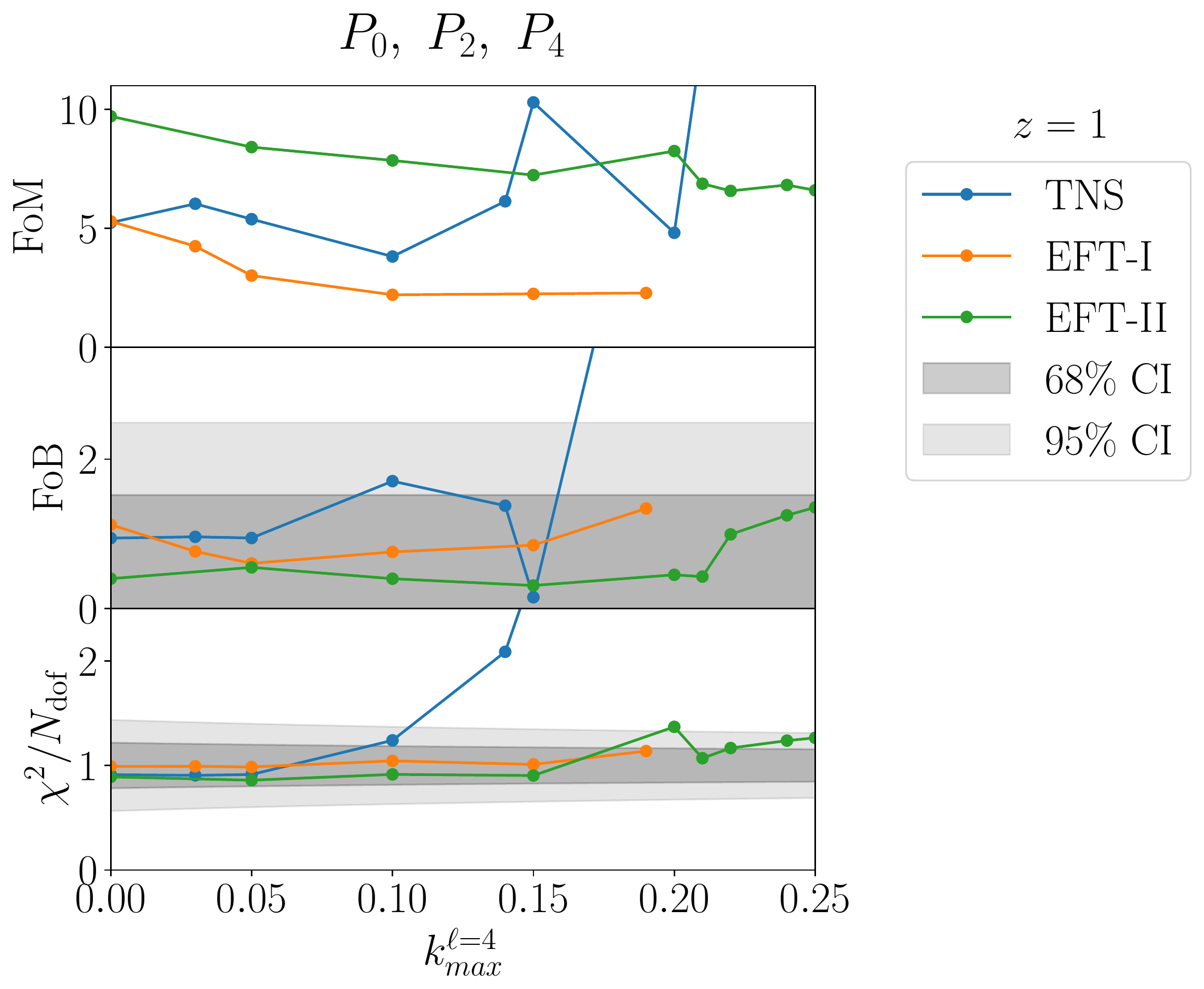}
    \caption{Performance measures for all three models for $z=1$. Left -- neglecting the hexadecapole, as a function of the maximum wave-number of the monopole and quadrupole. Right -- including the hexadecapole as a function of its maximum wave-number, with the maximum wave-numbers for monopole and quadrupole being $k_{\rm max}=0.22~h/{\rm Mpc}$ for the TNS model, $k_{\rm max}=0.19~h/{\rm Mpc}$ for EFT-I and $k_{\rm max}=0.3~h/{\rm Mpc}$ for EFT-II. Confidence intervals are computed in the way described in Fig.~\ref{All_perf_z05}.}
    \label{All_perf_z1}
\end{figure}

It is clear from the plots for the monopole and quadrupole analysis (left panels of Fig.~\ref{All_perf_z05} and Fig.~\ref{All_perf_z1}) that, as the bias increases beyond the 68\% confidence level, the FoM begins decreasing. This is due to a degeneracy between $w$ and $A$ that can be explained as follows. In a $w$CDM cosmology, linear growth is enhanced for $w<-1$, since matter domination ends later than in the corresponding $\Lambda$CDM cosmology, with the opposite happening for $w>-1$. The additional effect of the interaction is to add (subtract) effective friction for $A>0$ ($A<0$) which decreases (increases) growth. Therefore, given that the sign of $A$ is required to be the same as the sign of $(w+1)$ these effects act in the same direction, and the effect of a very negative $w$ can be mimicked by a very negative $A$ (and vice-versa). For this reason, these two effects are hard to distinguish and lead to an enhancement of the uncertainty for cases in which the posterior is sufficiently biased away from $w=-1$, $A=0$. This can be clearly seen in Fig.~\ref{MCMC_all_wA_bias}, in which the parameter bias is large. Similar effects have been previously found in other interacting dark energy models~\cite{2010JCAP...09..029L,Caldera-Cabral:2009hoy}.

Adding the hexadecapole, we see a marginal increase in the figure-of-merit for $k_{\rm max}^{\ell=4}=0.05~h/{\rm Mpc}$ in the TNS model. For both EFT-based models, the hexadecapole only acts to decrease the FoM, even though it does not bias the measurement substantially. The maximum values of $k_{\rm max}^{\ell=4}$ are quoted in Table~\ref{MostkmaxP4z05}.
\begin{table}[h]
\centering
 \begin{tabular}{||c c c c||} 
 \hline 
 Model &\ \ &$k_{\rm max}^{\ell=0,2}$ $(h/{\rm Mpc})$ &$k_{\rm max}^{\ell=4}$ $(h/{\rm Mpc})$\\
 \hline\hline
 TNS &\ \ & 0.22 & 0.10\\  
 \hline
 EFT-I &\ \ & 0.16 & 0.16\\ 
 \hline
 EFT-II &\ \ & 0.25 & 0.25\\ 
 \hline
\end{tabular}
\caption{Maximum values of $k_{\rm max}^{\ell=4}$ (with fixed $k_{\rm max}^{\ell=0,2}$) for which both the bias and the reduced chi-squared are within their 68\% confidence intervals, for each model for the analysis at $z=0.5$.}
\label{MostkmaxP4z05}
\end{table}
For the TNS model, the bias inverts sign after $k_{\rm max}^{\ell=4}=0.1~h/{\rm Mpc}$, as the hexadecapole data appears to prefer less growth than the monopole and quadrupole data. It therefore appears to be possible to find less biased results for $k_{\rm max}^{\ell=4}=0.14~h/{\rm Mpc}$, with a higher FoM. However, the price of this improvement is that the fit is worsened substantially, with the reduced chi-squared deviating far beyond its expected value due to the two types of data pushing the fit in different directions. That this phenomenon occurs in the TNS model but not in the EFT-based models is a result of the existence of the $O(\mu^4)$ counterterm in the latter. This parameter is being used to fit the hexadecapole data better and is unconstrained by the monopole and quadrupole. Additionally, it somewhat alleviates the pressure on the cosmological parameters, hence pushing stronger biases to larger $k_{\rm max}^{\ell=4}$ in the EFT models. In the TNS model, there are also $O(\mu^4)$ contributions from the $D_{\rm FoG}$ term, but these are constrained also by the monopole and quadrupole, since they depend on the single parameter $\sigma_v$.
Since it is clear that the modelling is failing for TNS at these larger values of $k_{\rm max}^{\ell=4}$, we consider these settings unacceptable.

\begin{figure}[h]
    \centering
		\includegraphics[width=0.6\textwidth]{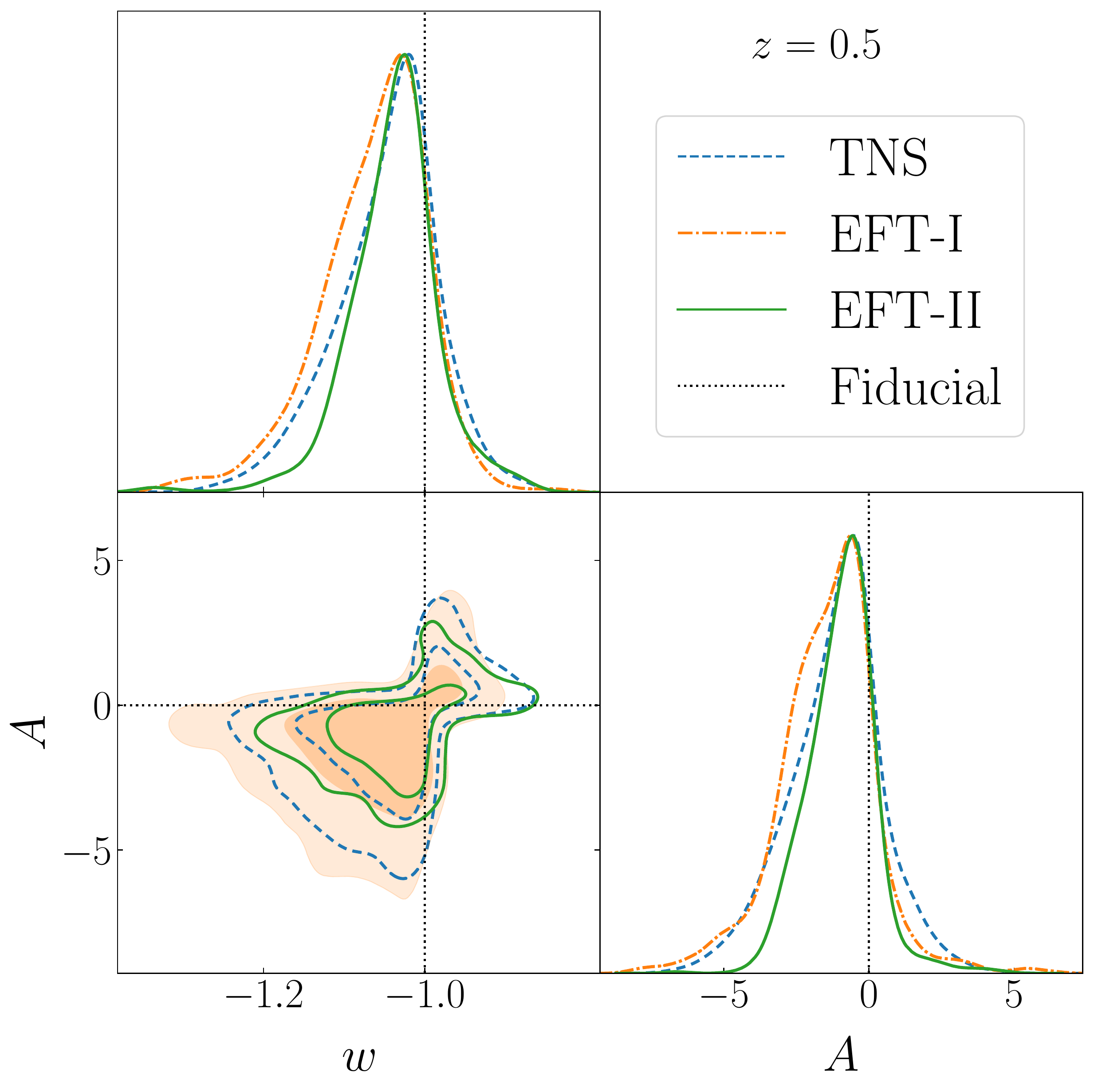}
    \caption{Marginal posterior distributions for the dark energy parameters in all models for $z=0.5$ in the best case scenarios of Table~\ref{MostFoM}. The fact that the contour shows non-vanishing values for regions forbidden by the prior ($A>0$, $w<-1$ and $A<0$, $w>-1$) is an artifact of the smoothing used for plotting, as the MCMC samples no points in those regions.}
    \label{MCMC_all_wA_z05}
\end{figure}

We show results for the posteriors of the dark energy parameters for all models at $z=0.5$ in the triangle plot of Fig.~\ref{MCMC_all_wA_z05} for the choices of $k_{\rm max}$ resulting in the tightest constraints and acceptable bias. Their details are given in Table~\ref{MostFoM}, in which we provide the FoM, the FoB as well as the means of the parameters along with their forecasted $1\sigma$ errors. Note that, in these and other contour plots, the 2D posterior for $w-A$ is substantially non-Gaussian, given the physical prior chosen for those parameters, i.e. ${\rm sign}(1+w)={\rm sign}(A)$. 

\begin{table}[h]
\centering
 \begin{tabular}{||c c c c c c c c||} 
 \hline 
 Model &\ \ &$k_{\rm max}^{\ell=0,2}$ $(h/{\rm Mpc})$& $k_{\rm max}^{\ell=4}$ $(h/{\rm Mpc})$ & FoM & FoB & $w$ & $A$ (b/GeV) \\
 \hline\hline
 TNS &\ \ & 0.22 & 0.05 & 9.62 & 0.85 & $-1.047^{+0.068}_{-0.052}$ & $-1.1^{+1.6}_{-1.5}$\\  
 \hline
 EFT-I &\ \ & 0.18 & -- & 8.22 & 1.12 & $-1.064^{+0.074}_{-0.049}$ & $-1.4^{+1.6}_{-1.3}$\\ 
 \hline
 EFT-II &\ \ & 0.27 & -- & 14.1 & 0.91 & $-1.039^{+0.059}_{-0.059}$ & $-0.9^{+1.2}_{-1.0}$\\ 
 \hline
\end{tabular}
\caption{Results for each model for the analysis at $z=0.5$ for the best cases -- obeying three conditions: i) maximal figure-of-merit; ii) FoB within the 68\% confidence interval range and iii) $\chi^2/N_{\rm dof}$ within the 68\% confidence interval range. An absent value for $k_{\rm max}^{\ell=4}$ means the hexadecapole is not included. The parameters $w$ and $A$ are presented as the mean and the limits of the 68\% confidence interval.}
\label{MostFoM}
\end{table}

It is clear from all these results that the EFT-II model is the best at describing the small scales at this redshift and is the one which allows for the retrieval of the most information about the parameters of this dark energy model from the data, resulting in the forecasted constraints below:
\be
w=-1.039\pm0.059\,,\ A=-0.9^{+1.2}_{-1.0}~{\rm b/GeV}\,.
\ee\\

For redshift $z=1$, the results for the performance measures are shown in Fig.~\ref{All_perf_z1}. This case is similar to  $z=0.5$, having  a smaller overall FoM. This is due to the fact that at this redshift the effects of dark energy are smaller than at lower redshifts and the power spectrum is therefore less sensitive to the parameters $w$ and $A$. Still, since non-linearities are not as developed at this higher redshift, there are slightly higher limiting values of $k_{\rm max}$ for the monopole and quadrupole, which can be seen in Table~\ref{Mostkmaxz1}.
\begin{table}[h]
\centering
 \begin{tabular}{||c c c||} 
 \hline 
 Model &\ \ &$k_{\rm max}^{\ell=0,2}$ $(h/{\rm Mpc})$\\
 \hline\hline
 TNS &\ \ & 0.23\\  
 \hline
 EFT-I &\ \ & 0.19\\ 
 \hline
 EFT-II &\ \ & >0.3\\ 
 \hline
\end{tabular}
\caption{Maximum values of $k_{\rm max}^{\ell=0,2}$ for which both the bias and the reduced chi-squared are within their 68\% confidence intervals, for each model for the analysis at $z=1$.}
\label{Mostkmaxz1}
\end{table}
For the EFT-II model, it is clear that the reach is likely to be larger than $k_{\rm max}^{\ell=0,2}=0.30~h/{\rm Mpc}$ and our tests suggest a maximum wave-number of $k_{\rm max}^{\ell=0,2}=0.35~h/{\rm Mpc}$. However, our PICOLA simulations are not expected to be sufficiently accurate above $k=0.30~h/{\rm Mpc}$ and we therefore prefer not to trust those results. This is the reason we chose not to plot them in Fig.~\ref{All_perf_z1}. Regardless of this, the EFT-II model achieves a FoM of $\sim10$, considerably higher than the other two models, which are again similar to each other when considering their best unbiased points.

Again here, when the hexadecapole is added, slight improvements of FoM are obtained for $k_{\rm max}^{\ell=4}\leq0.05~h/{\rm Mpc}$ for the TNS model. For this model, including more hexadecapole data either decreases the FoM or degrades the quality of the fit, similarly to the $z=0.5$ case. For both EFT-based models, the hexadecapole does not add any new information, even though it is still well described up to large values of $k_{\rm max}^{\ell=4}$, as can be seen in Table~\ref{MostkmaxP4z1}. We show full contour plots showing the effect of the hexadecapole for all models in  Figs.~\ref{MCMC_tns_compP4},~\ref{MCMC_eft_compP4} and \ref{MCMC_eft2_compP4} of Appendix~\ref{app:fullcount}.
\begin{table}[h]
\centering
 \begin{tabular}{||c c c c||} 
 \hline 
 Model &\ \ &$k_{\rm max}^{\ell=0,2}$  $(h/{\rm Mpc})$&$k_{\rm max}^{\ell=4}$ $(h/{\rm Mpc})$\\
 \hline\hline
 TNS &\ \ & 0.22 & 0.1\\  
 \hline
 EFT-I &\ \ & 0.19 & 0.19\\ 
 \hline
 EFT-II &\ \ & 0.3 & 0.2\\ 
 \hline
\end{tabular}
\caption{Maximum values of $k_{\rm max}^{\ell=4}$ for which both the bias and the reduced chi-squared are within their 68\% confidence intervals, for each model for the analysis at $z=1$. The fixed values of $k_{\rm max}^{\ell=0,2}$ are also shown and are chosen to be the best points from the analysis of the monopole and quadrupole for each model and therefore do not always correspond to those on Table~\ref{Mostkmaxz1}, since the later can have smaller FoM.}
\label{MostkmaxP4z1}
\end{table}

\begin{table}[h]
\centering
 \begin{tabular}{||c c c c c c c c||} 
 \hline 
 Model &\ \ &$k_{\rm max}^{\ell=0,2}$ $(h/{\rm Mpc})$& $k_{\rm max}^{\ell=4}$ $(h/{\rm Mpc})$ & FoM & FoB & $w$ & $A$ (b/GeV) \\
 \hline\hline
 TNS &\ \ & 0.22
& 0.03 & 6.03 & 0.96 & $-1.065^{+0.085}_{-0.057}$ & $-1.6^{+2.0}_{-1.8}$\\  
 \hline
 EFT-I &\ \ & 0.19 & -- & 5.29 & 1.12 & $-1.075^{+0.082}_{-0.050}$ & $-1.9^{+2.0}_{-1.6}$\\ 
 \hline
 EFT-II &\ \ & 0.3 & -- & 9.71 & 0.40 & $-1.022^{+0.070}_{-0.048}$ & $-0.0^{+1.7}_{-2.2}$\\ 
 \hline
\end{tabular}
\caption{Results for each model for the analysis at $z=1$ for the best cases -- obeying three conditions: i) maximal figure-of-merit; ii) FoB within the 68\% confidence interval range and iii) $\chi^2/N_{\rm dof}$ within the 68\% confidence interval range. An absent value for $k_{\rm max}^{\ell=4}$ means the hexadecapole is not included. The parameters $w$ and $A$ are presented as the mean and the limits of the 68\% confidence interval.}
\label{MostFoMz1}
\end{table}

\begin{figure}[h]
    \centering
		\includegraphics[width=0.6\textwidth]{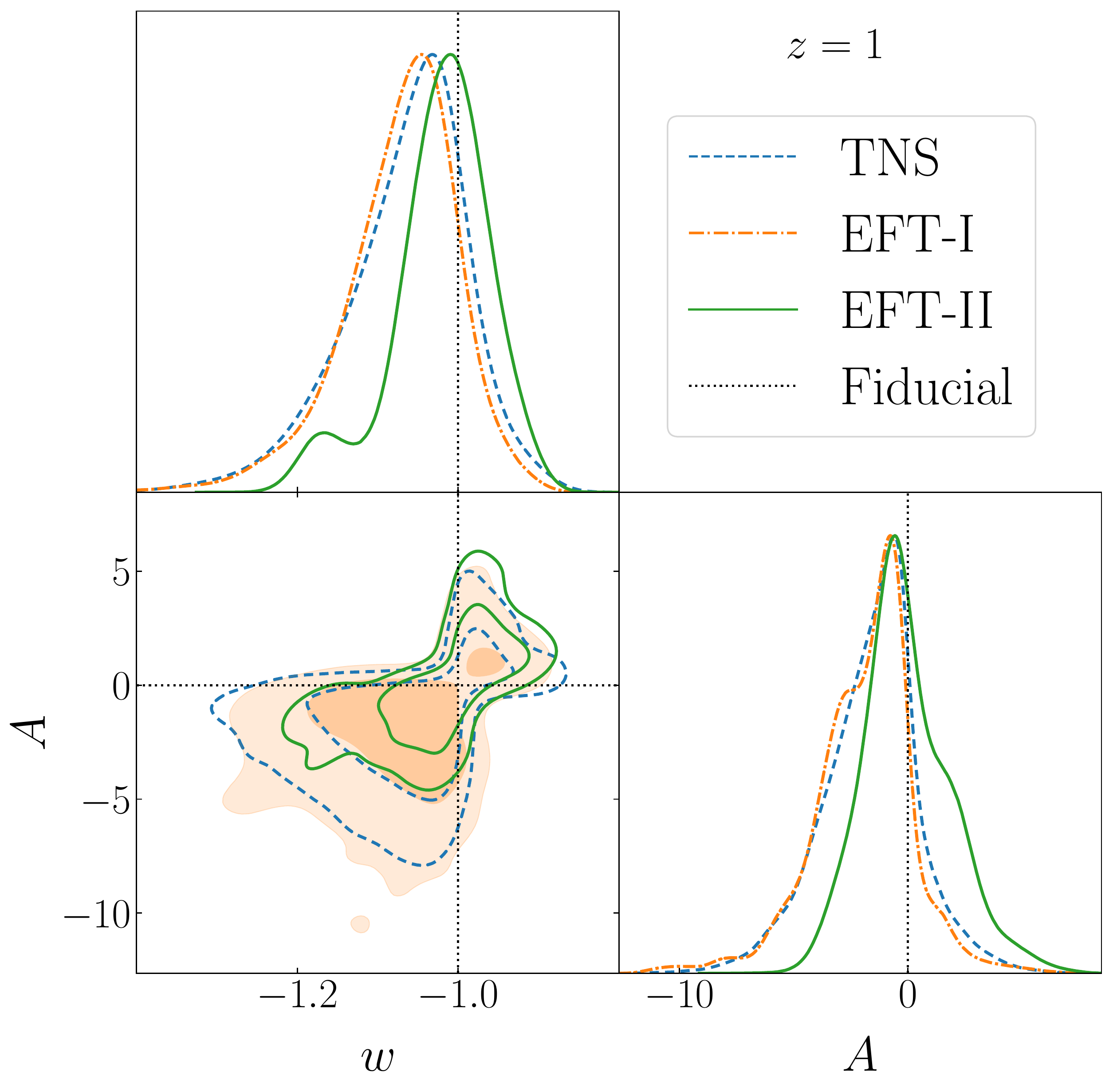}
    \caption{Marginal posterior distributions for the dark energy parameters in all models for $z=1$ in the best case scenarios of Table~\ref{MostFoMz1}.}
    \label{MCMC_all_wA_z1}
\end{figure}

\begin{figure}[h]
    \centering
		\includegraphics[width=0.49\textwidth]{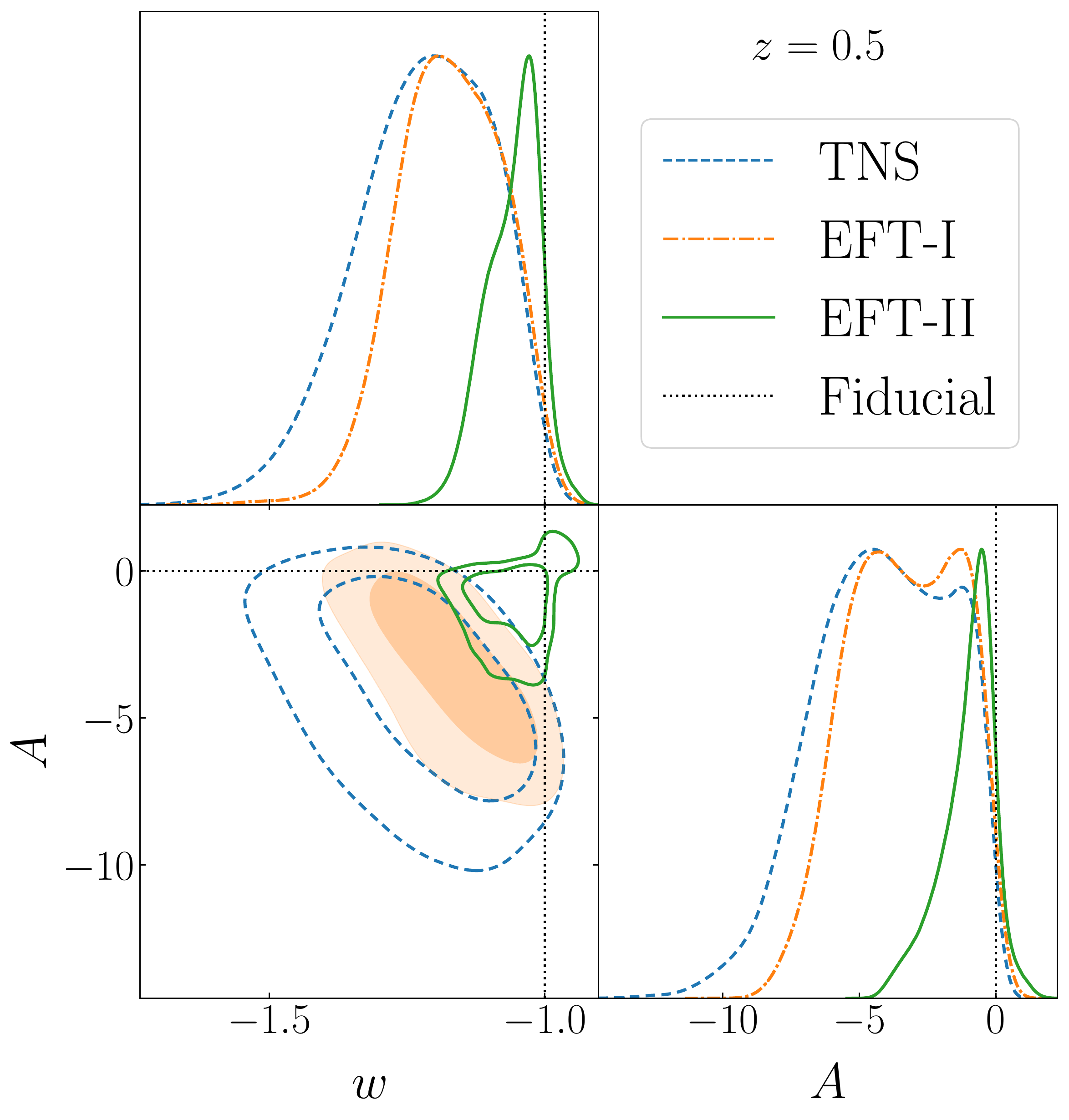}
		\includegraphics[width=0.49\textwidth]{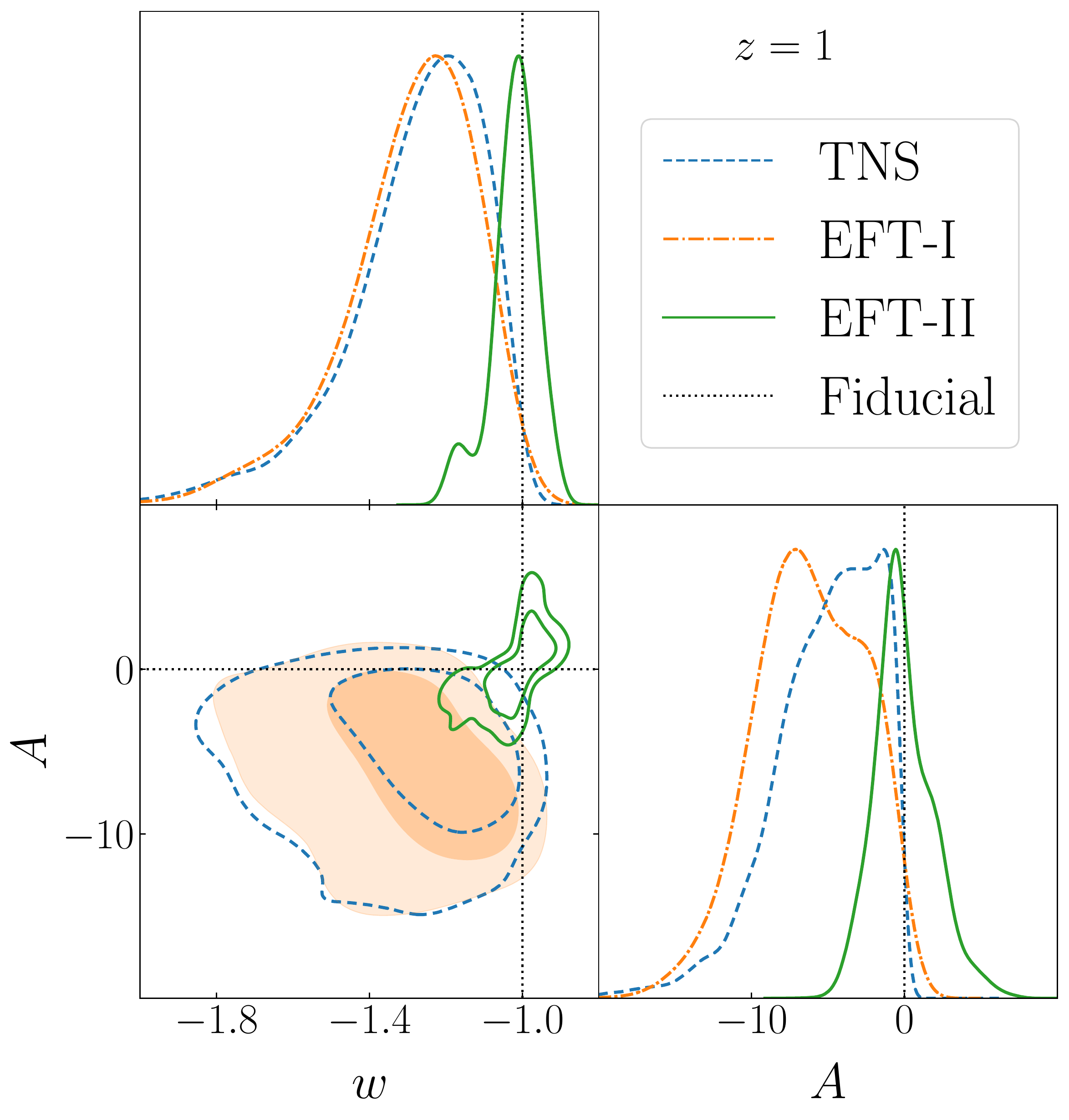}
    \caption{Marginal posterior distributions for the dark energy parameters in all models for $z=0.5$ (left) and $z=1$ (right) with $k_{\rm max}^{\ell=0,2}=0.3~h/{\rm Mpc}$ for all models. The hexadecapole is not included. Contrary to Figs.~\ref{MCMC_all_wA_z05} and \ref{MCMC_all_wA_z1}, here we include scales that are not accurately described by all models and thus bias the measurement of the dark energy parameters.} 
    \label{MCMC_all_wA_bias}
\end{figure}

The posteriors of the dark energy parameters, $w$ and $A$ for the analysis at $z=1$ can be seen in the triangle plot in Fig.~\ref{MCMC_all_wA_z1} for the choices of $k_{\rm max}$ leading to the best constraints in each model. Table~\ref{MostFoMz1} includes the detailed results for those scale choices. We can see once again that the EFT-II model performs best in all regards, with the TNS model doing slightly better than the EFT-I model. Our best marginalized forecasted errors at this redshift are, therefore,
\be
w=-1.022^{+0.070}_{-0.048}\,,\ A=-0.0^{+1.7}_{-2.2}~{\rm b/GeV}\,,
\ee
which is slightly worse than our best results for $z=0.5$.\\

The results described so far can also be used to assess the error made if the chosen value of $k_{\rm max}$ is outside the validity range of the model. This can be seen in Fig.~\ref{MCMC_all_wA_bias}, in which we show the forecasts for all models for $k_{\rm max}=0.3~h/{\rm Mpc}$. There it is clear that both the TNS and the EFT-I models incur in substantial biases, which would lead to a false detection of exotic dark energy effects. While the deviations from the expected values are greater for $z=1$, the uncertainties in the parameters are also larger in that case, which is why the figure-of-bias is generically larger for $z=0.5$, at which a stronger hint would be found for the wrong cosmology. A similar false detection can also arise in the interacting dark energy model studied in Ref.~\cite{DiValentino:2020leo}, albeit due to parameter degeneracies.

To better understand the reason behind the better performance of the EFT-II model, we run an additional test with different assumptions on the set of nuisance parameters. In this regard, there are two main differences between EFT-I and EFT-II. Firstly, the former assumes the local-Lagrangian relation on $b_{\gamma_2}$, Eq.~\eqref{lLb}, while the latter does not. Secondly, EFT-II includes an additional counterterm: $c_{\nabla^4 \delta}$ multiplying a term $\propto k^4 P_L(k)$. To assess the impact of these two extra parameters on the reach of the model we run four additional MCMC chains at fixed $k_{\rm max}^{\ell=0,2}=0.25~h/{\rm Mpc}$ and $k_{\rm max}^{\ell=4}=0.2~h/{\rm Mpc}$ at $z=0.5$, where EFT-II still gives a good fit but the FoB for EFT-I is already outside the $2\sigma$ confidence level. We adopt four different models:
\begin{itemize}
\item the baseline EFT-II, with $b_{\gamma_2}$ and $c_{\nabla^4 \delta}$ free;
\item a model with the local-Lagrangian relation $b_{\gamma_2}(b_1)$ and $c_{\nabla^4 \delta}$ free;
\item a model with $c_{\nabla^4 \delta}=0$ and $b_{\gamma_2}$ free;
\item a model with $b_{\gamma_2}(b_1)$ and $c_{\nabla^4 \delta}=0$, more similar to EFT-I.
\end{itemize}

\begin{figure}[h]
    \centering
		\includegraphics[width=0.65\textwidth]{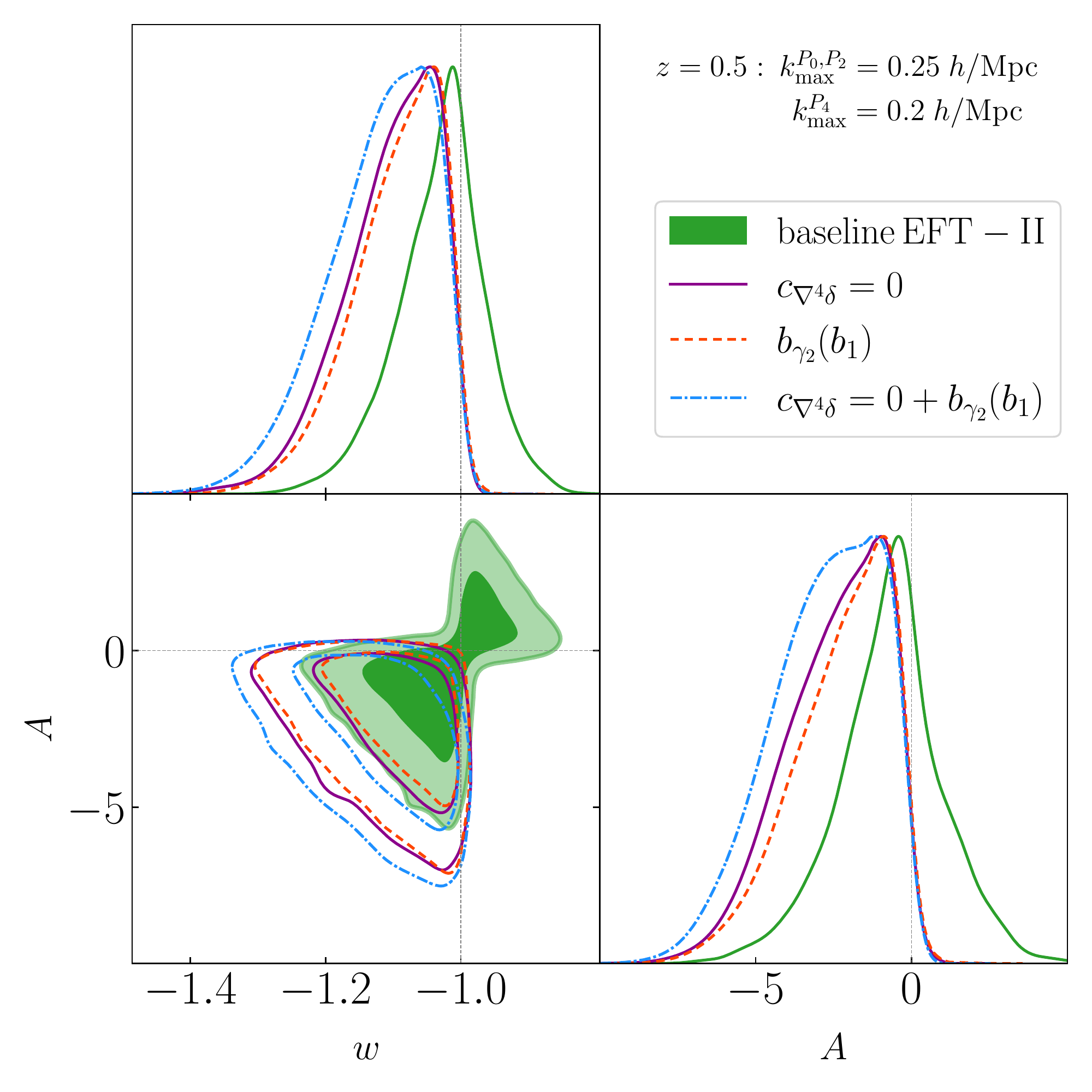}
    \caption{Marginal posterior distributions for the dark energy parameters in the tested versions of the EFT-II model for $z=0.5$ with $k_{\rm max}^{\ell=0,2}=0.25~h/{\rm Mpc}$ and $k_{\rm max}^{\ell=0,2}=0.2~h/{\rm Mpc}$. Filled green contours: baseline EFT-II model, with $\tilde{c}_{\nabla^4 \delta}$ anf $b_{\gamma_2}$ free, solid purple lines: model with $\tilde{c}_{\nabla^4 \delta}=0$, dashed red lines: model with the local-Lagrangian relation $b_{\gamma_2}(b_1)$, dot-dash blue lines: model with $\tilde{c}_{\nabla^4 \delta}=0$ and the local-Lagrangian relation.}
    \label{eft2-diff-bias-wA}
\end{figure}

The results are shown in Fig.~\ref{eft2-diff-bias-wA}, where we plot the 2D contours for $w-A$ marginalized over the other nuisance parameters. In can be seen that all configurations other than the baseline give biased results, similar to the ones shown in Fig.~\ref{MCMC_all_wA_bias} for the EFT-I model, albeit with smaller deviations from the fiducial values because of the lower $k_{\rm max}$ adopted for this test. It is thus clear that keeping both $b_{\gamma_2}$ and $c_{\nabla^4 \delta}$ as free parameters is key to increase the validity range of the EFT-II model. Moreover, this suggests that including both parameters in the EFT-I could extend its reach, making it similar to that of EFT-II. A full contour plot with all parameters for the four settings considered is shown in Fig.~\ref{eft2-diff-bias-full} of Appendix~\ref{app:fullcount}.

\section{Conclusions}\label{sec:conc}

We have studied the predictions of the dark scattering model for the redshift-space power spectrum of galaxies with the aim of forecasting the constraining power of stage-IV spectroscopic galaxy surveys for this model. We began by exploring different ways of modeling the dark matter power spectrum, the galaxy bias and the redshift-space prescription, arriving at three distinct models, which we labelled TNS, EFT-I and EFT-II. We then proceeded to compare them in terms of their ability to suitably extract the dark energy parameters, $w$ and $A$, without bias, before forecasting the uncertainties on those parameters achievable by stage-IV surveys.

In order to achieve this, we performed an MCMC analysis, based on PICOLA simulations of a $\Lambda$CDM cosmology at redshifts $z=0.5$ and $z=1$, with definitions mimicking DESI and Euclid-like settings, respectively. We find that the model labelled as EFT-II performs better than the other two models, as shown in Figs.~\ref{All_perf_z05} and \ref{All_perf_z1}, having both larger reach in the small scales and extracting more information from the data, achieving a FoM approximately 50\% higher than the other models. This is the most complex model, with the largest number of nuisance parameter (8), showing that for this particular analysis, complexity pays off. The other two models, TNS and EFT-I perform similarly well to each other, but describe the data accurately only up to smaller values of $k_{\rm max}$, when compared to EFT-II. These models therefore are not able to make optimal use of the range of data that will be available in stage-IV surveys and generate weaker constraints on the parameters of interest. In addition, using these models without taking their limitations into account can bias the measurement of $w$ and $A$ by over $2\sigma$, as seen in Fig.~\ref{MCMC_all_wA_bias}, possibly leading to false detection of interacting dark energy. These biases due to the insufficient modelling of non-linearities are not particular of exotic dark energy models and should also arise in any non-standard cosmology, as well as with other probes, as exemplified by the work of \cite{Schneider:2019xpf} focusing on weak lensing. It is therefore crucial to validate the modelling of small scales to avoid false detections and fictitious tensions appearing in future data analyses.

Similar analyses to this one had been performed, testing the reach of these models, such as \cite{Markovic:2019sva,Bose:2019psj,Bose:2019ywu}, comparing TNS and EFT-I. These works use the unbiased retrieval of the growth rate, $f$, as a probe of the performance instead of the model parameters, $w$ and $A$ used here. This choice is in effect a choice of the shape of the priors, as a flat prior on $f$ used in previous work is substantially different from the flat prior on $w$ and $A$ chosen here. This is likely the natural choice of the model building community, which is more interested in the details of fundamental parameters. We find similar results to \cite{Bose:2019ywu}  where TNS and EFT-I were compared - that TNS does slightly better than EFT-I. In addition to that difference, we do here a detailed comparison between EFT-II and the other two models, which had not yet been done in the literature.

Having compared all three perturbation theory models, we then produced forecasts for the dark energy parameters $w$ and $A$. Contour plots with the forecasts in the best case scenarios are shown in Fig.~\ref{MCMC_all_wA_z05} for $z=0.5$ and Fig.~\ref{MCMC_all_wA_z1} for $z=1$. For the best performing model, the forecasted errors for the two parameters of interest are $\sigma_w=0.06$, $\sigma_A=1.1$ b/GeV for the analysis at $z=0.5$ and $\sigma_w=0.06$, $\sigma_A=2$ b/GeV for $z=1$. These are the first detailed forecasts for the interaction parameter, $A$, and show the power that stage-IV surveys will have at constraining modifications to $\Lambda$CDM.

\section*{Acknowledgements}

PC and CM acknowledge support from a UK Research and Innovation Future Leaders Fellowship (MR/S016066/1). BB acknowledges support from the Swiss National
Science Foundation (SNSF) Professorship grant No. 170547. AP is a UK Research and Innovation Future Leaders Fellow, grant MR/S016066/1. 
This research utilised Queen Mary's Apocrita HPC facility, supported by QMUL Research-IT \url{http://doi.org/10.5281/zenodo.438045}. KM's part of the research was carried out at the Jet Propulsion Laboratory, California Institute of Technology, under a contract with the National Aeronautics and Space Administration (80NM0018D0004). We acknowledge the use of open source software \cite{scipy:2001,Hunter:2007,  mckinney-proc-scipy-2010, numpy:2011,  Lewis:1999bs, Lewis:2019xzd, McEwen:2016, Fang:2017}.

\bibliographystyle{JHEPmodplain}
\bibliography{IDE_biblio}

\providecommand{\href}[2]{#2}\begingroup\raggedright\begin{thebibliography}{100}

\bibitem{Abbott:2021bzy}
{\bf DES} Collaboration, T.~M.~C. Abbott {\em et~al.}, {\it {Dark Energy Survey
  Year 3 Results: Cosmological Constraints from Galaxy Clustering and Weak
  Lensing}},  \href{http://arxiv.org/abs/2105.13549}{{\sf arXiv:2105.13549}}.

\bibitem{Aghamousa:2016zmz}
{\bf DESI} Collaboration, A.~Aghamousa {\em et~al.}, {\it {The DESI Experiment
  Part I: Science,Targeting, and Survey Design}},
  \href{http://arxiv.org/abs/1611.00036}{{\sf arXiv:1611.00036}}.

\bibitem{Blanchard:2019oqi}
{\bf Euclid} Collaboration, A.~Blanchard {\em et~al.}, {\it {Euclid
  preparation: VII. Forecast validation for Euclid cosmological probes}},  {\sl
  Astron. Astrophys.} {\bf 642} (2020) A191,
  [\href{http://arxiv.org/abs/1910.09273}{{\sf arXiv:1910.09273}}],
  [\href{http://dx.doi.org/10.1051/0004-6361/202038071}{{\sf
  doi:10.1051/0004-6361/202038071}}].

\bibitem{Laureijs:2011gra}
{\bf EUCLID} Collaboration, R.~Laureijs {\em et~al.}, {\it {Euclid Definition
  Study Report}},  \href{http://arxiv.org/abs/1110.3193}{{\sf
  arXiv:1110.3193}}.

\bibitem{spergel2015wide}
D.~Spergel, N.~Gehrels, C.~Baltay, D.~Bennett, J.~Breckinridge, M.~Donahue,
  A.~Dressler, B.~Gaudi, T.~Greene, O.~Guyon, {\em et~al.}, {\it Wide-field
  infrarred survey telescope-astrophysics focused telescope assets wfirst-afta
  2015 report},  {\sl arXiv preprint arXiv:1503.03757} (2015)
  [\href{http://arxiv.org/abs/1503.03757}{{\sf arXiv:1503.03757}}].

\bibitem{Mandelbaum:2018ouv}
{\bf LSST Dark Energy Science} Collaboration, D.~Alonso {\em et~al.}, {\it {The
  LSST Dark Energy Science Collaboration (DESC) Science Requirements
  Document}},  \href{http://arxiv.org/abs/1809.01669}{{\sf arXiv:1809.01669}}.

\bibitem{Aghanim:2018eyx}
{\bf Planck} Collaboration, N.~Aghanim {\em et~al.}, {\it {Planck 2018 results.
  VI. Cosmological parameters}},  {\sl Astron. Astrophys.} {\bf 641} (2020) A6,
  [\href{http://arxiv.org/abs/1807.06209}{{\sf arXiv:1807.06209}}],
  [\href{http://dx.doi.org/10.1051/0004-6361/201833910}{{\sf
  doi:10.1051/0004-6361/201833910}}].

\bibitem{Anderson_2012}
L.~Anderson, E.~Aubourg, S.~Bailey, D.~Bizyaev, M.~Blanton, A.~S. Bolton,
  J.~Brinkmann, J.~R. Brownstein, A.~Burden, A.~J. Cuesta, and et~al., {\it The
  clustering of galaxies in the sdss-iii baryon oscillation spectroscopic
  survey: baryon acoustic oscillations in the data release 9 spectroscopic
  galaxy sample},  {\sl Monthly Notices of the Royal Astronomical Society} {\bf
  427} (Dec, 2012) 3435–3467,
  [\href{http://dx.doi.org/10.1111/j.1365-2966.2012.22066.x}{{\sf
  doi:10.1111/j.1365-2966.2012.22066.x}}].

\bibitem{Song:2015oza}
Y.-S. Song, A.~Taruya, E.~Linder, K.~Koyama, C.~G. Sabiu, G.-B. Zhao,
  F.~Bernardeau, T.~Nishimichi, and T.~Okumura, {\it {Consistent Modified
  Gravity Analysis of Anisotropic Galaxy Clustering Using BOSS DR11}},  {\sl
  Phys. Rev. D} {\bf 92} (2015), no.~4 043522,
  [\href{http://arxiv.org/abs/1507.01592}{{\sf arXiv:1507.01592}}],
  [\href{http://dx.doi.org/10.1103/PhysRevD.92.043522}{{\sf
  doi:10.1103/PhysRevD.92.043522}}].

\bibitem{Beutler_2016}
F.~Beutler, H.-J. Seo, S.~Saito, C.-H. Chuang, A.~J. Cuesta, D.~J. Eisenstein,
  H.~Gil-Marín, J.~N. Grieb, N.~Hand, F.-S. Kitaura, and et~al., {\it The
  clustering of galaxies in the completed sdss-iii baryon oscillation
  spectroscopic survey: anisotropic galaxy clustering in fourier space},  {\sl
  Monthly Notices of the Royal Astronomical Society} {\bf 466} (Dec, 2016)
  2242–2260, [\href{http://dx.doi.org/10.1093/mnras/stw3298}{{\sf
  doi:10.1093/mnras/stw3298}}].

\bibitem{SpurioMancini:2019rxy}
A.~Spurio~Mancini, F.~K\"ohlinger, B.~Joachimi, V.~Pettorino, B.~M. Sch\"afer,
  R.~Reischke, E.~van Uitert, S.~Brieden, M.~Archidiacono, and J.~Lesgourgues,
  {\it {KiDS + GAMA: constraints on horndeski gravity from combined large-scale
  structure probes}},  {\sl Mon. Not. Roy. Astron. Soc.} {\bf 490} (2019),
  no.~2 2155--2177, [\href{http://arxiv.org/abs/1901.03686}{{\sf
  arXiv:1901.03686}}], [\href{http://dx.doi.org/10.1093/mnras/stz2581}{{\sf
  doi:10.1093/mnras/stz2581}}].

\bibitem{Troster:2019ean}
T.~Tr\"oster {\em et~al.}, {\it {Cosmology from large-scale structure:
  Constraining $\Lambda$CDM with BOSS}},  {\sl Astron. Astrophys.} {\bf 633}
  (2020) L10, [\href{http://arxiv.org/abs/1909.11006}{{\sf arXiv:1909.11006}}],
  [\href{http://dx.doi.org/10.1051/0004-6361/201936772}{{\sf
  doi:10.1051/0004-6361/201936772}}].

\bibitem{Alam:2020sor}
{\bf eBOSS} Collaboration, S.~Alam {\em et~al.}, {\it {Completed SDSS-IV
  extended Baryon Oscillation Spectroscopic Survey: Cosmological implications
  from two decades of spectroscopic surveys at the Apache Point Observatory}},
  {\sl Phys. Rev. D} {\bf 103} (2021), no.~8 083533,
  [\href{http://arxiv.org/abs/2007.08991}{{\sf arXiv:2007.08991}}],
  [\href{http://dx.doi.org/10.1103/PhysRevD.103.083533}{{\sf
  doi:10.1103/PhysRevD.103.083533}}].

\bibitem{Troster:2020kai}
{\bf KiDS} Collaboration, T.~Tr\"oster {\em et~al.}, {\it {KiDS-1000 Cosmology:
  constraints beyond flat $\Lambda$CDM}},  {\sl Astron. Astrophys.} {\bf 649}
  (2021) A88, [\href{http://arxiv.org/abs/2010.16416}{{\sf arXiv:2010.16416}}],
  [\href{http://dx.doi.org/10.1051/0004-6361/202039805}{{\sf
  doi:10.1051/0004-6361/202039805}}].

\bibitem{Caldwell:1997ii}
R.~R. Caldwell, R.~Dave, and P.~J. Steinhardt, {\it {Cosmological imprint of an
  energy component with general equation of state}},  {\sl Phys. Rev. Lett.}
  {\bf 80} (1998) 1582--1585,
  [\href{http://arxiv.org/abs/astro-ph/9708069}{{\sf arXiv:astro-ph/9708069}}],
  [\href{http://dx.doi.org/10.1103/PhysRevLett.80.1582}{{\sf
  doi:10.1103/PhysRevLett.80.1582}}].

\bibitem{Amendola:1999er}
L.~Amendola, {\it {Coupled quintessence}},  {\sl Phys. Rev. D} {\bf 62} (2000)
  043511, [\href{http://arxiv.org/abs/astro-ph/9908023}{{\sf
  arXiv:astro-ph/9908023}}],
  [\href{http://dx.doi.org/10.1103/PhysRevD.62.043511}{{\sf
  doi:10.1103/PhysRevD.62.043511}}].

\bibitem{Peebles:2002gy}
P.~J.~E. Peebles and B.~Ratra, {\it {The Cosmological Constant and Dark
  Energy}},  {\sl Rev. Mod. Phys.} {\bf 75} (2003) 559--606,
  [\href{http://arxiv.org/abs/astro-ph/0207347}{{\sf arXiv:astro-ph/0207347}}],
  [\href{http://dx.doi.org/10.1103/RevModPhys.75.559}{{\sf
  doi:10.1103/RevModPhys.75.559}}].

\bibitem{Copeland:2006wr}
E.~J. Copeland, M.~Sami, and S.~Tsujikawa, {\it {Dynamics of dark energy}},
  {\sl Int. J. Mod. Phys. D} {\bf 15} (2006) 1753--1936,
  [\href{http://arxiv.org/abs/hep-th/0603057}{{\sf arXiv:hep-th/0603057}}],
  [\href{http://dx.doi.org/10.1142/S021827180600942X}{{\sf
  doi:10.1142/S021827180600942X}}].

\bibitem{Nojiri:2006ri}
S.~Nojiri and S.~D. Odintsov, {\it {Introduction to modified gravity and
  gravitational alternative for dark energy}},  {\sl eConf} {\bf C0602061}
  (2006) 06, [\href{http://arxiv.org/abs/hep-th/0601213}{{\sf
  arXiv:hep-th/0601213}}],
  [\href{http://dx.doi.org/10.1142/S0219887807001928}{{\sf
  doi:10.1142/S0219887807001928}}].

\bibitem{Sotiriou:2008rp}
T.~P. Sotiriou and V.~Faraoni, {\it {f(R) Theories Of Gravity}},  {\sl Rev.
  Mod. Phys.} {\bf 82} (2010) 451--497,
  [\href{http://arxiv.org/abs/0805.1726}{{\sf arXiv:0805.1726}}],
  [\href{http://dx.doi.org/10.1103/RevModPhys.82.451}{{\sf
  doi:10.1103/RevModPhys.82.451}}].

\bibitem{DeFelice:2010aj}
A.~De~Felice and S.~Tsujikawa, {\it {f(R) theories}},  {\sl Living Rev. Rel.}
  {\bf 13} (2010) 3, [\href{http://arxiv.org/abs/1002.4928}{{\sf
  arXiv:1002.4928}}], [\href{http://dx.doi.org/10.12942/lrr-2010-3}{{\sf
  doi:10.12942/lrr-2010-3}}].

\bibitem{Clifton:2011jh}
T.~Clifton, P.~G. Ferreira, A.~Padilla, and C.~Skordis, {\it {Modified Gravity
  and Cosmology}},  {\sl Phys. Rept.} {\bf 513} (2012) 1--189,
  [\href{http://arxiv.org/abs/1106.2476}{{\sf arXiv:1106.2476}}],
  [\href{http://dx.doi.org/10.1016/j.physrep.2012.01.001}{{\sf
  doi:10.1016/j.physrep.2012.01.001}}].

\bibitem{Bertolami:2012xn}
O.~Bertolami, P.~Carrilho, and J.~Paramos, {\it {Two-scalar-field model for the
  interaction of dark energy and dark matter}},  {\sl Phys. Rev. D} {\bf 86}
  (2012) 103522, [\href{http://arxiv.org/abs/1206.2589}{{\sf
  arXiv:1206.2589}}], [\href{http://dx.doi.org/10.1103/PhysRevD.86.103522}{{\sf
  doi:10.1103/PhysRevD.86.103522}}].

\bibitem{Pourtsidou:2013nha}
A.~Pourtsidou, C.~Skordis, and E.~J. Copeland, {\it {Models of dark matter
  coupled to dark energy}},  {\sl Phys. Rev. D} {\bf 88} (2013), no.~8 083505,
  [\href{http://arxiv.org/abs/1307.0458}{{\sf arXiv:1307.0458}}],
  [\href{http://dx.doi.org/10.1103/PhysRevD.88.083505}{{\sf
  doi:10.1103/PhysRevD.88.083505}}].

\bibitem{Guzzo:2008ac}
L.~Guzzo {\em et~al.}, {\it {A test of the nature of cosmic acceleration using
  galaxy redshift distortions}},  {\sl Nature} {\bf 451} (2008) 541--545,
  [\href{http://arxiv.org/abs/0802.1944}{{\sf arXiv:0802.1944}}],
  [\href{http://dx.doi.org/10.1038/nature06555}{{\sf
  doi:10.1038/nature06555}}].

\bibitem{Blake:2011rj}
C.~Blake {\em et~al.}, {\it {The WiggleZ Dark Energy Survey: the growth rate of
  cosmic structure since redshift z=0.9}},  {\sl Mon. Not. Roy. Astron. Soc.}
  {\bf 415} (2011) 2876, [\href{http://arxiv.org/abs/1104.2948}{{\sf
  arXiv:1104.2948}}],
  [\href{http://dx.doi.org/10.1111/j.1365-2966.2011.18903.x}{{\sf
  doi:10.1111/j.1365-2966.2011.18903.x}}].

\bibitem{Reid:2012sw}
B.~A. Reid {\em et~al.}, {\it {The clustering of galaxies in the SDSS-III
  Baryon Oscillation Spectroscopic Survey: measurements of the growth of
  structure and expansion rate at z=0.57 from anisotropic clustering}},  {\sl
  Mon. Not. Roy. Astron. Soc.} {\bf 426} (2012) 2719,
  [\href{http://arxiv.org/abs/1203.6641}{{\sf arXiv:1203.6641}}],
  [\href{http://dx.doi.org/10.1111/j.1365-2966.2012.21779.x}{{\sf
  doi:10.1111/j.1365-2966.2012.21779.x}}].

\bibitem{Beutler:2013yhm}
{\bf BOSS} Collaboration, F.~Beutler {\em et~al.}, {\it {The clustering of
  galaxies in the SDSS-III Baryon Oscillation Spectroscopic Survey: Testing
  gravity with redshift-space distortions using the power spectrum
  multipoles}},  {\sl Mon. Not. Roy. Astron. Soc.} {\bf 443} (2014), no.~2
  1065--1089, [\href{http://arxiv.org/abs/1312.4611}{{\sf arXiv:1312.4611}}],
  [\href{http://dx.doi.org/10.1093/mnras/stu1051}{{\sf
  doi:10.1093/mnras/stu1051}}].

\bibitem{Macaulay_2013}
E.~Macaulay, I.~K. Wehus, and H.~K. Eriksen, {\it Lower growth rate from recent
  redshift space distortion measurements than expected from planck},  {\sl
  Physical Review Letters} {\bf 111} (Oct, 2013)
  [\href{http://dx.doi.org/10.1103/physrevlett.111.161301}{{\sf
  doi:10.1103/physrevlett.111.161301}}].

\bibitem{Gil-Marin:2016wya}
H.~Gil-Mar\'\i{}n, W.~J. Percival, L.~Verde, J.~R. Brownstein, C.-H. Chuang,
  F.-S. Kitaura, S.~A. Rodr\'\i{}guez-Torres, and M.~D. Olmstead, {\it {The
  clustering of galaxies in the SDSS-III Baryon Oscillation Spectroscopic
  Survey: RSD measurement from the power spectrum and bispectrum of the DR12
  BOSS galaxies}},  {\sl Mon. Not. Roy. Astron. Soc.} {\bf 465} (2017), no.~2
  1757--1788, [\href{http://arxiv.org/abs/1606.00439}{{\sf arXiv:1606.00439}}],
  [\href{http://dx.doi.org/10.1093/mnras/stw2679}{{\sf
  doi:10.1093/mnras/stw2679}}].

\bibitem{Chisari:2019tus}
N.~E. Chisari {\em et~al.}, {\it {Modelling baryonic feedback for survey
  cosmology}},  {\sl Open J. Astrophys.} {\bf 2} (2019), no.~1 4,
  [\href{http://arxiv.org/abs/1905.06082}{{\sf arXiv:1905.06082}}],
  [\href{http://dx.doi.org/10.21105/astro.1905.06082}{{\sf
  doi:10.21105/astro.1905.06082}}].

\bibitem{Markovic:2019sva}
K.~Markovic, B.~Bose, and A.~Pourtsidou, {\it {Assessing non-linear models for
  galaxy clustering I: unbiased growth forecasts from multipole expansion}},
  {\sl Open J. Astrophys.} {\bf 2} (2019), no.~1 13,
  [\href{http://arxiv.org/abs/1904.11448}{{\sf arXiv:1904.11448}}],
  [\href{http://dx.doi.org/10.21105/astro.1904.11448}{{\sf
  doi:10.21105/astro.1904.11448}}].

\bibitem{Bose:2019psj}
B.~Bose, A.~Pourtsidou, K.~Markovi\v{c}, and F.~Beutler, {\it {Assessing
  non-linear models for galaxy clustering II: model validation and forecasts
  for Stage IV surveys}},  \href{http://arxiv.org/abs/1905.05122}{{\sf
  arXiv:1905.05122}}, \href{http://dx.doi.org/10.1093/mnras/staa502}{{\sf
  doi:10.1093/mnras/staa502}}.

\bibitem{Schneider:2019snl}
A.~Schneider, N.~Stoira, A.~Refregier, A.~J. Weiss, M.~Knabenhans, J.~Stadel,
  and R.~Teyssier, {\it {Baryonic effects for weak lensing. Part I. Power
  spectrum and covariance matrix}},  {\sl JCAP} {\bf 04} (2020) 019,
  [\href{http://arxiv.org/abs/1910.11357}{{\sf arXiv:1910.11357}}],
  [\href{http://dx.doi.org/10.1088/1475-7516/2020/04/019}{{\sf
  doi:10.1088/1475-7516/2020/04/019}}].

\bibitem{Nishimichi:2020tvu}
T.~Nishimichi, G.~D'Amico, M.~M. Ivanov, L.~Senatore, M.~Simonovi\'c,
  M.~Takada, M.~Zaldarriaga, and P.~Zhang, {\it {Blinded challenge for
  precision cosmology with large-scale structure: results from effective field
  theory for the redshift-space galaxy power spectrum}},  {\sl Phys. Rev. D}
  {\bf 102} (2020), no.~12 123541, [\href{http://arxiv.org/abs/2003.08277}{{\sf
  arXiv:2003.08277}}],
  [\href{http://dx.doi.org/10.1103/PhysRevD.102.123541}{{\sf
  doi:10.1103/PhysRevD.102.123541}}].

\bibitem{Martinelli:2020yto}
{\bf Euclid} Collaboration, M.~Martinelli {\em et~al.}, {\it {Euclid: impact of
  nonlinear prescriptions on cosmological parameter estimation from weak
  lensing cosmic shear}},  \href{http://arxiv.org/abs/2010.12382}{{\sf
  arXiv:2010.12382}}.

\bibitem{Pezzotta:2021vfn}
A.~Pezzotta, M.~Crocce, A.~Eggemeier, A.~G. S\'anchez, and R.~Scoccimarro, {\it
  {Testing one-loop galaxy bias: cosmological constraints from the power
  spectrum}},  \href{http://arxiv.org/abs/2102.08315}{{\sf arXiv:2102.08315}}.

\bibitem{Secco:2021vhm}
{\bf DES} Collaboration, L.~F. Secco {\em et~al.}, {\it {Dark Energy Survey
  Year 3 Results: Cosmology from Cosmic Shear and Robustness to Modeling
  Uncertainty}},  \href{http://arxiv.org/abs/2105.13544}{{\sf
  arXiv:2105.13544}}.

\bibitem{Taruya:2010mx}
A.~Taruya, T.~Nishimichi, and S.~Saito, {\it {Baryon Acoustic Oscillations in
  2D: Modeling Redshift-space Power Spectrum from Perturbation Theory}},  {\sl
  Phys.Rev.} {\bf D82} (2010) 063522,
  [\href{http://arxiv.org/abs/1006.0699}{{\sf arXiv:1006.0699}}],
  [\href{http://dx.doi.org/10.1103/PhysRevD.82.063522}{{\sf
  doi:10.1103/PhysRevD.82.063522}}].

\bibitem{Baumann:2010tm}
D.~Baumann, A.~Nicolis, L.~Senatore, and M.~Zaldarriaga, {\it {Cosmological
  Non-Linearities as an Effective Fluid}},  {\sl JCAP} {\bf 07} (2012) 051,
  [\href{http://arxiv.org/abs/1004.2488}{{\sf arXiv:1004.2488}}],
  [\href{http://dx.doi.org/10.1088/1475-7516/2012/07/051}{{\sf
  doi:10.1088/1475-7516/2012/07/051}}].

\bibitem{Carrasco:2012cv}
J.~J.~M. Carrasco, M.~P. Hertzberg, and L.~Senatore, {\it {The Effective Field
  Theory of Cosmological Large Scale Structures}},  {\sl JHEP} {\bf 09} (2012)
  082, [\href{http://arxiv.org/abs/1206.2926}{{\sf arXiv:1206.2926}}],
  [\href{http://dx.doi.org/10.1007/JHEP09(2012)082}{{\sf
  doi:10.1007/JHEP09(2012)082}}].

\bibitem{DAmico:2019fhj}
G.~D'Amico, J.~Gleyzes, N.~Kokron, K.~Markovic, L.~Senatore, P.~Zhang,
  F.~Beutler, and H.~Gil-Mar\'\i{}n, {\it {The Cosmological Analysis of the
  SDSS/BOSS data from the Effective Field Theory of Large-Scale Structure}},
  {\sl JCAP} {\bf 05} (2020) 005, [\href{http://arxiv.org/abs/1909.05271}{{\sf
  arXiv:1909.05271}}],
  [\href{http://dx.doi.org/10.1088/1475-7516/2020/05/005}{{\sf
  doi:10.1088/1475-7516/2020/05/005}}].

\bibitem{Ivanov:2019pdj}
M.~M. Ivanov, M.~Simonovi\'c, and M.~Zaldarriaga, {\it {Cosmological Parameters
  from the BOSS Galaxy Power Spectrum}},  {\sl JCAP} {\bf 05} (2020) 042,
  [\href{http://arxiv.org/abs/1909.05277}{{\sf arXiv:1909.05277}}],
  [\href{http://dx.doi.org/10.1088/1475-7516/2020/05/042}{{\sf
  doi:10.1088/1475-7516/2020/05/042}}].

\bibitem{2020arXiv200811284D}
E.~{Di Valentino} {\em et~al.}, {\it {Cosmology Intertwined II: The Hubble
  Constant Tension}},  {\sl arXiv e-prints} (Aug., 2020) arXiv:2008.11284,
  [\href{http://arxiv.org/abs/2008.11284}{{\sf arXiv:2008.11284}}].

\bibitem{DiValentino:2020vvd}
E.~Di~Valentino {\em et~al.}, {\it {Cosmology Intertwined III: $f \sigma_8$ and
  $S_8$}},  \href{http://arxiv.org/abs/2008.11285}{{\sf arXiv:2008.11285}}.

\bibitem{2021APh...13102607D}
E.~{Di Valentino} {\em et~al.}, {\it {Cosmology Intertwined IV: The age of the
  universe and its curvature}},  {\sl Astroparticle Physics} {\bf 131} (Sept.,
  2021) 102607, [\href{http://arxiv.org/abs/2008.11286}{{\sf
  arXiv:2008.11286}}],
  [\href{http://dx.doi.org/10.1016/j.astropartphys.2021.102607}{{\sf
  doi:10.1016/j.astropartphys.2021.102607}}].

\bibitem{Verde:2019ivm}
L.~Verde, T.~Treu, and A.~G. Riess, {\it {Tensions between the Early and the
  Late Universe}},  {\sl Nature Astron.} {\bf 3} (7, 2019) 891,
  [\href{http://arxiv.org/abs/1907.10625}{{\sf arXiv:1907.10625}}],
  [\href{http://dx.doi.org/10.1038/s41550-019-0902-0}{{\sf
  doi:10.1038/s41550-019-0902-0}}].

\bibitem{Knox:2019rjx}
L.~Knox and M.~Millea, {\it {Hubble constant hunter\textquoteright{}s guide}},
  {\sl Phys. Rev. D} {\bf 101} (2020), no.~4 043533,
  [\href{http://arxiv.org/abs/1908.03663}{{\sf arXiv:1908.03663}}],
  [\href{http://dx.doi.org/10.1103/PhysRevD.101.043533}{{\sf
  doi:10.1103/PhysRevD.101.043533}}].

\bibitem{Jedamzik_2021}
K.~Jedamzik, L.~Pogosian, and G.-B. Zhao, {\it {Why reducing the cosmic sound
  horizon alone can not fully resolve the Hubble tension}},  {\sl
  Communications Physics} {\bf 4} (Jun, 2021)
  [\href{http://dx.doi.org/10.1038/s42005-021-00628-x}{{\sf
  doi:10.1038/s42005-021-00628-x}}].

\bibitem{Di_Valentino_2021}
E.~Di~Valentino, O.~Mena, S.~Pan, L.~Visinelli, W.~Yang, A.~Melchiorri, D.~F.
  Mota, A.~G. Riess, and J.~Silk, {\it {In the realm of the Hubble tension—a
  review of solutions}},  {\sl Classical and Quantum Gravity} {\bf 38} (Jul,
  2021) 153001, [\href{http://dx.doi.org/10.1088/1361-6382/ac086d}{{\sf
  doi:10.1088/1361-6382/ac086d}}].

\bibitem{perivolaropoulos2021challenges}
L.~Perivolaropoulos and F.~Skara, {\it {Challenges for $\Lambda$CDM: An
  update}},  2021.

\bibitem{Abbott:2020knk}
{\bf DES} Collaboration, T.~M.~C. Abbott {\em et~al.}, {\it {Dark Energy Survey
  Year 1 Results: Cosmological constraints from cluster abundances and weak
  lensing}},  {\sl Phys. Rev. D} {\bf 102} (2020), no.~2 023509,
  [\href{http://arxiv.org/abs/2002.11124}{{\sf arXiv:2002.11124}}],
  [\href{http://dx.doi.org/10.1103/PhysRevD.102.023509}{{\sf
  doi:10.1103/PhysRevD.102.023509}}].

\bibitem{2021A&A...646A.140H}
C.~Heymans {\em et~al.}, {\it {KiDS-1000 Cosmology: Multi-probe weak
  gravitational lensing and spectroscopic galaxy clustering constraints}},
  {\sl Astron. Astrophys.} {\bf 646} (2021) A140,
  [\href{http://arxiv.org/abs/2007.15632}{{\sf arXiv:2007.15632}}],
  [\href{http://dx.doi.org/10.1051/0004-6361/202039063}{{\sf
  doi:10.1051/0004-6361/202039063}}].

\bibitem{Vikhlinin:2008ym}
A.~Vikhlinin {\em et~al.}, {\it {Chandra Cluster Cosmology Project III:
  Cosmological Parameter Constraints}},  {\sl Astrophys. J.} {\bf 692} (2009)
  1060--1074, [\href{http://arxiv.org/abs/0812.2720}{{\sf arXiv:0812.2720}}],
  [\href{http://dx.doi.org/10.1088/0004-637X/692/2/1060}{{\sf
  doi:10.1088/0004-637X/692/2/1060}}].

\bibitem{deHaan:2016qvy}
{\bf SPT} Collaboration, T.~de~Haan {\em et~al.}, {\it {Cosmological
  Constraints from Galaxy Clusters in the 2500 square-degree SPT-SZ Survey}},
  {\sl Astrophys. J.} {\bf 832} (2016), no.~1 95,
  [\href{http://arxiv.org/abs/1603.06522}{{\sf arXiv:1603.06522}}],
  [\href{http://dx.doi.org/10.3847/0004-637X/832/1/95}{{\sf
  doi:10.3847/0004-637X/832/1/95}}].

\bibitem{Simpson:2015yfa}
F.~Simpson, C.~Blake, J.~A. Peacock, I.~Baldry, J.~Bland-Hawthorn, A.~Heavens,
  C.~Heymans, J.~Loveday, and P.~Norberg, {\it {Galaxy and mass assembly:
  Redshift space distortions from the clipped galaxy field}},  {\sl Phys. Rev.
  D} {\bf 93} (2016), no.~2 023525,
  [\href{http://arxiv.org/abs/1505.03865}{{\sf arXiv:1505.03865}}],
  [\href{http://dx.doi.org/10.1103/PhysRevD.93.023525}{{\sf
  doi:10.1103/PhysRevD.93.023525}}].

\bibitem{Bean_2008}
R.~Bean, E.~E. Flanagan, I.~Laszlo, and M.~Trodden, {\it Constraining
  interactions in cosmology’s dark sector},  {\sl Physical Review D} {\bf 78}
  (Dec, 2008) [\href{http://dx.doi.org/10.1103/physrevd.78.123514}{{\sf
  doi:10.1103/physrevd.78.123514}}].

\bibitem{Xia:2009zzb}
J.-Q. Xia, {\it {Constraint on coupled dark energy models from observations}},
  {\sl Phys. Rev. D} {\bf 80} (2009) 103514,
  [\href{http://arxiv.org/abs/0911.4820}{{\sf arXiv:0911.4820}}],
  [\href{http://dx.doi.org/10.1103/PhysRevD.80.103514}{{\sf
  doi:10.1103/PhysRevD.80.103514}}].

\bibitem{Amendola_2012}
L.~Amendola, V.~Pettorino, C.~Quercellini, and A.~Vollmer, {\it Testing coupled
  dark energy with next-generation large-scale observations},  {\sl Physical
  Review D} {\bf 85} (May, 2012)
  [\href{http://dx.doi.org/10.1103/physrevd.85.103008}{{\sf
  doi:10.1103/physrevd.85.103008}}].

\bibitem{Gomez-Valent:2020mqn}
A.~G\'omez-Valent, V.~Pettorino, and L.~Amendola, {\it {Update on coupled dark
  energy and the $H_0$ tension}},  {\sl Phys. Rev. D} {\bf 101} (2020), no.~12
  123513, [\href{http://arxiv.org/abs/2004.00610}{{\sf arXiv:2004.00610}}],
  [\href{http://dx.doi.org/10.1103/PhysRevD.101.123513}{{\sf
  doi:10.1103/PhysRevD.101.123513}}].

\bibitem{Pourtsidou:2016ico}
A.~Pourtsidou and T.~Tram, {\it {Reconciling CMB and structure growth
  measurements with dark energy interactions}},  {\sl Phys. Rev. D} {\bf 94}
  (2016), no.~4 043518, [\href{http://arxiv.org/abs/1604.04222}{{\sf
  arXiv:1604.04222}}],
  [\href{http://dx.doi.org/10.1103/PhysRevD.94.043518}{{\sf
  doi:10.1103/PhysRevD.94.043518}}].

\bibitem{Skordis:2015yra}
C.~Skordis, A.~Pourtsidou, and E.~J. Copeland, {\it {Parametrized
  post-Friedmannian framework for interacting dark energy theories}},  {\sl
  Phys. Rev. D} {\bf 91} (2015), no.~8 083537,
  [\href{http://arxiv.org/abs/1502.07297}{{\sf arXiv:1502.07297}}],
  [\href{http://dx.doi.org/10.1103/PhysRevD.91.083537}{{\sf
  doi:10.1103/PhysRevD.91.083537}}].

\bibitem{Richarte:2014yva}
M.~G. Richarte and L.~Xu, {\it {Interacting parametrized post-Friedmann
  method}},  {\sl Gen. Rel. Grav.} {\bf 48} (2016), no.~4 39,
  [\href{http://arxiv.org/abs/1407.4348}{{\sf arXiv:1407.4348}}],
  [\href{http://dx.doi.org/10.1007/s10714-016-2035-4}{{\sf
  doi:10.1007/s10714-016-2035-4}}].

\bibitem{Simpson:2010vh}
F.~Simpson, {\it {Scattering of dark matter and dark energy}},  {\sl Phys. Rev.
  D} {\bf 82} (2010) 083505, [\href{http://arxiv.org/abs/1007.1034}{{\sf
  arXiv:1007.1034}}], [\href{http://dx.doi.org/10.1103/PhysRevD.82.083505}{{\sf
  doi:10.1103/PhysRevD.82.083505}}].

\bibitem{Baldi:2016zom}
M.~Baldi and F.~Simpson, {\it {Structure formation simulations with momentum
  exchange: alleviating tensions between high-redshift and low-redshift
  cosmological probes}},  {\sl Mon. Not. Roy. Astron. Soc.} {\bf 465} (2017),
  no.~1 653--666, [\href{http://arxiv.org/abs/1605.05623}{{\sf
  arXiv:1605.05623}}], [\href{http://dx.doi.org/10.1093/mnras/stw2702}{{\sf
  doi:10.1093/mnras/stw2702}}].

\bibitem{Lesgourgues:2015wza}
J.~Lesgourgues, G.~Marques-Tavares, and M.~Schmaltz, {\it {Evidence for dark
  matter interactions in cosmological precision data?}},  {\sl JCAP} {\bf 02}
  (2016) 037, [\href{http://arxiv.org/abs/1507.04351}{{\sf arXiv:1507.04351}}],
  [\href{http://dx.doi.org/10.1088/1475-7516/2016/02/037}{{\sf
  doi:10.1088/1475-7516/2016/02/037}}].

\bibitem{Linton:2017ged}
M.~S. Linton, A.~Pourtsidou, R.~Crittenden, and R.~Maartens, {\it {Variable
  sound speed in interacting dark energy models}},  {\sl JCAP} {\bf 04} (2018)
  043, [\href{http://arxiv.org/abs/1711.05196}{{\sf arXiv:1711.05196}}],
  [\href{http://dx.doi.org/10.1088/1475-7516/2018/04/043}{{\sf
  doi:10.1088/1475-7516/2018/04/043}}].

\bibitem{Bose:2017jjx}
B.~Bose, M.~Baldi, and A.~Pourtsidou, {\it {Modelling Non-Linear Effects of
  Dark Energy}},  {\sl JCAP} {\bf 04} (2018) 032,
  [\href{http://arxiv.org/abs/1711.10976}{{\sf arXiv:1711.10976}}],
  [\href{http://dx.doi.org/10.1088/1475-7516/2018/04/032}{{\sf
  doi:10.1088/1475-7516/2018/04/032}}].

\bibitem{Buen-Abad:2017gxg}
M.~A. Buen-Abad, M.~Schmaltz, J.~Lesgourgues, and T.~Brinckmann, {\it
  {Interacting Dark Sector and Precision Cosmology}},  {\sl JCAP} {\bf 01}
  (2018) 008, [\href{http://arxiv.org/abs/1708.09406}{{\sf arXiv:1708.09406}}],
  [\href{http://dx.doi.org/10.1088/1475-7516/2018/01/008}{{\sf
  doi:10.1088/1475-7516/2018/01/008}}].

\bibitem{Kase:2019mox}
R.~Kase and S.~Tsujikawa, {\it {Weak cosmic growth in coupled dark energy with
  a Lagrangian formulation}},  {\sl Phys. Lett. B} {\bf 804} (2020) 135400,
  [\href{http://arxiv.org/abs/1911.02179}{{\sf arXiv:1911.02179}}],
  [\href{http://dx.doi.org/10.1016/j.physletb.2020.135400}{{\sf
  doi:10.1016/j.physletb.2020.135400}}].

\bibitem{Chamings:2019kcl}
F.~N. Chamings, A.~Avgoustidis, E.~J. Copeland, A.~M. Green, and A.~Pourtsidou,
  {\it {Understanding the suppression of structure formation from dark
  matter-dark energy momentum coupling}},  {\sl Phys. Rev. D} {\bf 101} (2020),
  no.~4 043531, [\href{http://arxiv.org/abs/1912.09858}{{\sf
  arXiv:1912.09858}}],
  [\href{http://dx.doi.org/10.1103/PhysRevD.101.043531}{{\sf
  doi:10.1103/PhysRevD.101.043531}}].

\bibitem{Amendola:2020ldb}
L.~Amendola and S.~Tsujikawa, {\it {Scaling solutions and weak gravity in dark
  energy with energy and momentum couplings}},  {\sl JCAP} {\bf 06} (2020) 020,
  [\href{http://arxiv.org/abs/2003.02686}{{\sf arXiv:2003.02686}}],
  [\href{http://dx.doi.org/10.1088/1475-7516/2020/06/020}{{\sf
  doi:10.1088/1475-7516/2020/06/020}}].

\bibitem{1869592}
J.~B. Jim\'enez, D.~Bettoni, D.~Figueruelo, F.~A.~T. Pannia, and S.~Tsujikawa,
  {\it {Probing elastic interactions in the dark sector and the role of
  $S_8$}},  \href{http://arxiv.org/abs/2106.11222}{{\sf arXiv:2106.11222}}.

\bibitem{Baldi:2014ica}
M.~Baldi and F.~Simpson, {\it {Simulating Momentum Exchange in the Dark
  Sector}},  {\sl Mon. Not. Roy. Astron. Soc.} {\bf 449} (2015), no.~3
  2239--2249, [\href{http://arxiv.org/abs/1412.1080}{{\sf arXiv:1412.1080}}],
  [\href{http://dx.doi.org/10.1093/mnras/stv405}{{\sf
  doi:10.1093/mnras/stv405}}].

\bibitem{Malik:2008im}
K.~A. Malik and D.~Wands, {\it {Cosmological perturbations}},  {\sl Phys.
  Rept.} {\bf 475} (2009) 1--51, [\href{http://arxiv.org/abs/0809.4944}{{\sf
  arXiv:0809.4944}}],
  [\href{http://dx.doi.org/10.1016/j.physrep.2009.03.001}{{\sf
  doi:10.1016/j.physrep.2009.03.001}}].

\bibitem{Jain:1994}
B.~{Jain} and E.~{Bertschinger}, {\it {Second-Order Power Spectrum and
  Nonlinear Evolution at High Redshift}},  {\sl Astrophys. J.} {\bf 431} (Aug,
  1994) 495, [\href{http://arxiv.org/abs/astro-ph/9311070}{{\sf
  arXiv:astro-ph/9311070}}], [\href{http://dx.doi.org/10.1086/174502}{{\sf
  doi:10.1086/174502}}].

\bibitem{Bernardeau:2001qr}
F.~Bernardeau, S.~Colombi, E.~Gaztanaga, and R.~Scoccimarro, {\it {Large scale
  structure of the universe and cosmological perturbation theory}},  {\sl Phys.
  Rept.} {\bf 367} (2002) 1--248,
  [\href{http://arxiv.org/abs/astro-ph/0112551}{{\sf arXiv:astro-ph/0112551}}],
  [\href{http://dx.doi.org/10.1016/S0370-1573(02)00135-7}{{\sf
  doi:10.1016/S0370-1573(02)00135-7}}].

\bibitem{McDonald:2009dh}
P.~McDonald and A.~Roy, {\it {Clustering of dark matter tracers: generalizing
  bias for the coming era of precision LSS}},  {\sl JCAP} {\bf 0908} (2009)
  020, [\href{http://arxiv.org/abs/0902.0991}{{\sf arXiv:0902.0991}}],
  [\href{http://dx.doi.org/10.1088/1475-7516/2009/08/020}{{\sf
  doi:10.1088/1475-7516/2009/08/020}}].

\bibitem{Assassi_2014}
V.~Assassi, D.~Baumann, D.~Green, and M.~Zaldarriaga, {\it Renormalized halo
  bias},  {\sl Journal of Cosmology and Astroparticle Physics} {\bf 2014} (Aug,
  2014) 056–056, [\href{http://dx.doi.org/10.1088/1475-7516/2014/08/056}{{\sf
  doi:10.1088/1475-7516/2014/08/056}}].

\bibitem{Desjacques:2016bnm}
V.~Desjacques, D.~Jeong, and F.~Schmidt, {\it {Large-Scale Galaxy Bias}},  {\sl
  Phys. Rept.} {\bf 733} (2018) 1--193,
  [\href{http://arxiv.org/abs/1611.09787}{{\sf arXiv:1611.09787}}],
  [\href{http://dx.doi.org/10.1016/j.physrep.2017.12.002}{{\sf
  doi:10.1016/j.physrep.2017.12.002}}].

\bibitem{Fujita:2020}
T.~{Fujita}, V.~{Mauerhofer}, L.~{Senatore}, Z.~{Vlah}, and R.~{Angulo}, {\it
  {Very massive tracers and higher derivative biases}},  {\sl JCAP} {\bf 2020}
  (Jan., 2020) 009, [\href{http://arxiv.org/abs/1609.00717}{{\sf
  arXiv:1609.00717}}],
  [\href{http://dx.doi.org/10.1088/1475-7516/2020/01/009}{{\sf
  doi:10.1088/1475-7516/2020/01/009}}].

\bibitem{Chan:2012jj}
K.~C. Chan, R.~Scoccimarro, and R.~K. Sheth, {\it {Gravity and Large-Scale
  Non-local Bias}},  {\sl Phys. Rev.} {\bf D85} (2012) 083509,
  [\href{http://arxiv.org/abs/1201.3614}{{\sf arXiv:1201.3614}}],
  [\href{http://dx.doi.org/10.1103/PhysRevD.85.083509}{{\sf
  doi:10.1103/PhysRevD.85.083509}}].

\bibitem{Baldauf:2012hs}
T.~Baldauf, U.~Seljak, V.~Desjacques, and P.~McDonald, {\it {Evidence for
  Quadratic Tidal Tensor Bias from the Halo Bispectrum}},  {\sl Phys. Rev.}
  {\bf D86} (2012) 083540, [\href{http://arxiv.org/abs/1201.4827}{{\sf
  arXiv:1201.4827}}], [\href{http://dx.doi.org/10.1103/PhysRevD.86.083540}{{\sf
  doi:10.1103/PhysRevD.86.083540}}].

\bibitem{Sheth:2012fc}
R.~K. Sheth, K.~C. Chan, and R.~Scoccimarro, {\it {Nonlocal Lagrangian bias}},
  {\sl Phys. Rev.} {\bf D87} (2013), no.~8 083002,
  [\href{http://arxiv.org/abs/1207.7117}{{\sf arXiv:1207.7117}}],
  [\href{http://dx.doi.org/10.1103/PhysRevD.87.083002}{{\sf
  doi:10.1103/PhysRevD.87.083002}}].

\bibitem{Saito:2014qha}
S.~Saito, T.~Baldauf, Z.~Vlah, U.~Seljak, T.~Okumura, and P.~McDonald, {\it
  {Understanding higher-order nonlocal halo bias at large scales by combining
  the power spectrum with the bispectrum}},  {\sl Phys. Rev.} {\bf D90} (2014),
  no.~12 123522, [\href{http://arxiv.org/abs/1405.1447}{{\sf
  arXiv:1405.1447}}], [\href{http://dx.doi.org/10.1103/PhysRevD.90.123522}{{\sf
  doi:10.1103/PhysRevD.90.123522}}].

\bibitem{delaBella:2017qjy}
L.~F. de~la Bella, D.~Regan, D.~Seery, and S.~Hotchkiss, {\it {The matter power
  spectrum in redshift space using effective field theory}},  {\sl JCAP} {\bf
  11} (2017) 039, [\href{http://arxiv.org/abs/1704.05309}{{\sf
  arXiv:1704.05309}}],
  [\href{http://dx.doi.org/10.1088/1475-7516/2017/11/039}{{\sf
  doi:10.1088/1475-7516/2017/11/039}}].

\bibitem{Chudaykin:2020}
A.~Chudaykin, M.~M. Ivanov, O.~H.~E. Philcox, and M.~Simonovi\'c, {\it
  {Nonlinear perturbation theory extension of the Boltzmann code CLASS}},  {\sl
  Phys. Rev. D} {\bf 102} (2020), no.~6 063533,
  [\href{http://arxiv.org/abs/2004.10607}{{\sf arXiv:2004.10607}}],
  [\href{http://dx.doi.org/10.1103/PhysRevD.102.063533}{{\sf
  doi:10.1103/PhysRevD.102.063533}}].

\bibitem{Perko:2016puo}
A.~Perko, L.~Senatore, E.~Jennings, and R.~H. Wechsler, {\it {Biased Tracers in
  Redshift Space in the EFT of Large-Scale Structure}},
  \href{http://arxiv.org/abs/1610.09321}{{\sf arXiv:1610.09321}}.

\bibitem{Vlah:2015zda}
Z.~Vlah, U.~Seljak, M.~Y. Chu, and Y.~Feng, {\it {Perturbation theory,
  effective field theory, and oscillations in the power spectrum}},  {\sl JCAP}
  {\bf 03} (2016) 057, [\href{http://arxiv.org/abs/1509.02120}{{\sf
  arXiv:1509.02120}}],
  [\href{http://dx.doi.org/10.1088/1475-7516/2016/03/057}{{\sf
  doi:10.1088/1475-7516/2016/03/057}}].

\bibitem{Eisenstein:1997ik}
D.~J. Eisenstein and W.~Hu, {\it {Baryonic features in the matter transfer
  function}},  {\sl Astrophys. J.} {\bf 496} (1998) 605,
  [\href{http://arxiv.org/abs/astro-ph/9709112}{{\sf arXiv:astro-ph/9709112}}],
  [\href{http://dx.doi.org/10.1086/305424}{{\sf doi:10.1086/305424}}].

\bibitem{Schmittfull:2018yuk}
M.~Schmittfull, M.~Simonovi\'c, V.~Assassi, and M.~Zaldarriaga, {\it {Modeling
  Biased Tracers at the Field Level}},  {\sl Phys. Rev. D} {\bf 100} (2019),
  no.~4 043514, [\href{http://arxiv.org/abs/1811.10640}{{\sf
  arXiv:1811.10640}}],
  [\href{http://dx.doi.org/10.1103/PhysRevD.100.043514}{{\sf
  doi:10.1103/PhysRevD.100.043514}}].

\bibitem{Scoccimarro:1999}
R.~{Scoccimarro}, H.~M.~P. {Couchman}, and J.~A. {Frieman}, {\it {The
  Bispectrum as a Signature of Gravitational Instability in Redshift Space}},
  {\sl Astrophys. J.} {\bf 517} (June, 1999) 531--540,
  [\href{http://arxiv.org/abs/astro-ph/9808305}{{\sf arXiv:astro-ph/9808305}}],
  [\href{http://dx.doi.org/10.1086/307220}{{\sf doi:10.1086/307220}}].

\bibitem{McEwen:2016}
J.~E. {McEwen}, X.~{Fang}, C.~M. {Hirata}, and J.~A. {Blazek}, {\it {FAST-PT: a
  novel algorithm to calculate convolution integrals in cosmological
  perturbation theory}},  {\sl JCAP} {\bf 2016} (Sept., 2016) 015,
  [\href{http://arxiv.org/abs/1603.04826}{{\sf arXiv:1603.04826}}],
  [\href{http://dx.doi.org/10.1088/1475-7516/2016/09/015}{{\sf
  doi:10.1088/1475-7516/2016/09/015}}].

\bibitem{Fang:2017}
X.~{Fang}, J.~A. {Blazek}, J.~E. {McEwen}, and C.~M. {Hirata}, {\it {FAST-PT
  II: an algorithm to calculate convolution integrals of general tensor
  quantities in cosmological perturbation theory}},  {\sl JCAP} {\bf 2017}
  (Feb., 2017) 030, [\href{http://arxiv.org/abs/1609.05978}{{\sf
  arXiv:1609.05978}}],
  [\href{http://dx.doi.org/10.1088/1475-7516/2017/02/030}{{\sf
  doi:10.1088/1475-7516/2017/02/030}}].

\bibitem{Bose:2019ywu}
B.~Bose, K.~Koyama, and H.~A. Winther, {\it {Assessing non-linear models for
  galaxy clustering III: Theoretical accuracy for Stage IV surveys}},  {\sl
  JCAP} {\bf 10} (2019) 021, [\href{http://arxiv.org/abs/1905.05135}{{\sf
  arXiv:1905.05135}}],
  [\href{http://dx.doi.org/10.1088/1475-7516/2019/10/021}{{\sf
  doi:10.1088/1475-7516/2019/10/021}}].

\bibitem{delaBella:2018fdb}
L.~Fonseca de~la Bella, D.~Regan, D.~Seery, and D.~Parkinson, {\it {Impact of
  bias and redshift-space modelling for the halo power spectrum: Testing the
  effective field theory of large-scale structure}},  {\sl JCAP} {\bf 07}
  (2020) 011, [\href{http://arxiv.org/abs/1805.12394}{{\sf arXiv:1805.12394}}],
  [\href{http://dx.doi.org/10.1088/1475-7516/2020/07/011}{{\sf
  doi:10.1088/1475-7516/2020/07/011}}].

\bibitem{Yang:2018euj}
W.~Yang, S.~Pan, E.~Di~Valentino, R.~C. Nunes, S.~Vagnozzi, and D.~F. Mota,
  {\it {Tale of stable interacting dark energy, observational signatures, and
  the $H_0$ tension}},  {\sl JCAP} {\bf 09} (2018) 019,
  [\href{http://arxiv.org/abs/1805.08252}{{\sf arXiv:1805.08252}}],
  [\href{http://dx.doi.org/10.1088/1475-7516/2018/09/019}{{\sf
  doi:10.1088/1475-7516/2018/09/019}}].

\bibitem{Howlett:2015hfa}
C.~Howlett, M.~Manera, and W.~J. Percival, {\it {L-PICOLA: A parallel code for
  fast dark matter simulation}},  {\sl Astron. Comput.} {\bf 12} (2015)
  109--126, [\href{http://arxiv.org/abs/1506.03737}{{\sf arXiv:1506.03737}}],
  [\href{http://dx.doi.org/10.1016/j.ascom.2015.07.003}{{\sf
  doi:10.1016/j.ascom.2015.07.003}}].

\bibitem{Winther:2017jof}
H.~A. Winther, K.~Koyama, M.~Manera, B.~S. Wright, and G.-B. Zhao, {\it {COLA
  with scale-dependent growth: applications to screened modified gravity
  models}},  {\sl JCAP} {\bf 1708} (2017), no.~08 006,
  [\href{http://arxiv.org/abs/1703.00879}{{\sf arXiv:1703.00879}}],
  [\href{http://dx.doi.org/10.1088/1475-7516/2017/08/006}{{\sf
  doi:10.1088/1475-7516/2017/08/006}}].

\bibitem{Hinshaw:2012aka}
{\bf WMAP} Collaboration, G.~Hinshaw {\em et~al.}, {\it {Nine-Year Wilkinson
  Microwave Anisotropy Probe (WMAP) Observations: Cosmological Parameter
  Results}},  {\sl Astrophys. J. Suppl.} {\bf 208} (2013) 19,
  [\href{http://arxiv.org/abs/1212.5226}{{\sf arXiv:1212.5226}}],
  [\href{http://dx.doi.org/10.1088/0067-0049/208/2/19}{{\sf
  doi:10.1088/0067-0049/208/2/19}}].

\bibitem{Foreman_Mackey_2013}
D.~Foreman-Mackey, D.~W. Hogg, D.~Lang, and J.~Goodman, {\it emcee: The mcmc
  hammer},  {\sl Publications of the Astronomical Society of the Pacific} {\bf
  125} (Mar, 2013) 306–312, [\href{http://dx.doi.org/10.1086/670067}{{\sf
  doi:10.1086/670067}}].

\bibitem{Osato_2019}
K.~Osato, T.~Nishimichi, F.~Bernardeau, and A.~Taruya, {\it Perturbation theory
  challenge for cosmological parameters estimation: Matter power spectrum in
  real space},  {\sl Physical Review D} {\bf 99} (Mar, 2019)
  [\href{http://dx.doi.org/10.1103/physrevd.99.063530}{{\sf
  doi:10.1103/physrevd.99.063530}}].

\bibitem{Eggemeier:2020umu}
A.~Eggemeier, R.~Scoccimarro, M.~Crocce, A.~Pezzotta, and A.~G. S\'anchez, {\it
  {Testing one-loop galaxy bias: Power spectrum}},  {\sl Phys. Rev. D} {\bf
  102} (2020), no.~10 103530, [\href{http://arxiv.org/abs/2006.09729}{{\sf
  arXiv:2006.09729}}],
  [\href{http://dx.doi.org/10.1103/PhysRevD.102.103530}{{\sf
  doi:10.1103/PhysRevD.102.103530}}].

\bibitem{2010JCAP...09..029L}
L.~{Lopez Honorez}, B.~A. {Reid}, O.~{Mena}, L.~{Verde}, and R.~{Jimenez}, {\it
  {Coupled dark matter-dark energy in light of near universe observations}},
  {\sl {JCAP}} {\bf 2010} (Sept., 2010) 029,
  [\href{http://arxiv.org/abs/1006.0877}{{\sf arXiv:1006.0877}}],
  [\href{http://dx.doi.org/10.1088/1475-7516/2010/09/029}{{\sf
  doi:10.1088/1475-7516/2010/09/029}}].

\bibitem{Caldera-Cabral:2009hoy}
G.~Caldera-Cabral, R.~Maartens, and B.~M. Schaefer, {\it {The Growth of
  Structure in Interacting Dark Energy Models}},  {\sl JCAP} {\bf 07} (2009)
  027, [\href{http://arxiv.org/abs/0905.0492}{{\sf arXiv:0905.0492}}],
  [\href{http://dx.doi.org/10.1088/1475-7516/2009/07/027}{{\sf
  doi:10.1088/1475-7516/2009/07/027}}].

\bibitem{DiValentino:2020leo}
E.~Di~Valentino and O.~Mena, {\it {A fake Interacting Dark Energy detection?}},
   {\sl Mon. Not. Roy. Astron. Soc.} {\bf 500} (2020), no.~1 L22--L26,
  [\href{http://arxiv.org/abs/2009.12620}{{\sf arXiv:2009.12620}}],
  [\href{http://dx.doi.org/10.1093/mnrasl/slaa175}{{\sf
  doi:10.1093/mnrasl/slaa175}}].

\bibitem{Schneider:2019xpf}
A.~Schneider, A.~Refregier, S.~Grandis, D.~Eckert, N.~Stoira, T.~Kacprzak,
  M.~Knabenhans, J.~Stadel, and R.~Teyssier, {\it {Baryonic effects for weak
  lensing. Part II. Combination with X-ray data and extended cosmologies}},
  {\sl JCAP} {\bf 04} (2020) 020, [\href{http://arxiv.org/abs/1911.08494}{{\sf
  arXiv:1911.08494}}],
  [\href{http://dx.doi.org/10.1088/1475-7516/2020/04/020}{{\sf
  doi:10.1088/1475-7516/2020/04/020}}].

\bibitem{scipy:2001}
E.~Jones, T.~Oliphant, P.~Peterson, {\em et~al.}, {\it {SciPy}: Open source
  scientific tools for {Python}},  2001--.

\bibitem{Hunter:2007}
J.~D. Hunter, {\it Matplotlib: A 2d graphics environment},  {\sl Computing In
  Science \& Engineering} {\bf 9} (2007), no.~3 90--95.

\bibitem{mckinney-proc-scipy-2010}
W.~McKinney, {\it Data structures for statistical computing in python},  in
  {\em Proceedings of the 9th Python in Science Conference} (S.~van~der Walt
  and J.~Millman, eds.), pp.~51 -- 56, 2010.

\bibitem{numpy:2011}
S.~{Van Der Walt}, S.~C. {Colbert}, and G.~{Varoquaux}, {\it {The NumPy array:
  a structure for efficient numerical computation}},  {\sl ArXiv e-prints}
  (Feb., 2011) [\href{http://arxiv.org/abs/1102.1523}{{\sf arXiv:1102.1523}}].

\bibitem{Lewis:1999bs}
A.~Lewis, A.~Challinor, and A.~Lasenby, {\it {Efficient computation of CMB
  anisotropies in closed FRW models}},  {\sl Astrophys. J.} {\bf 538} (2000)
  473--476, [\href{http://arxiv.org/abs/astro-ph/9911177}{{\sf
  arXiv:astro-ph/9911177}}], [\href{http://dx.doi.org/10.1086/309179}{{\sf
  doi:10.1086/309179}}].

\bibitem{Lewis:2019xzd}
A.~Lewis, {\it {GetDist: a Python package for analysing Monte Carlo samples}},
  \href{http://arxiv.org/abs/1910.13970}{{\sf arXiv:1910.13970}}.

\end{thebibliography}\endgroup

\appendix

\section{Full contours}\label{app:fullcount}

\begin{figure}[h]
    \centering
		\includegraphics[width=\textwidth]{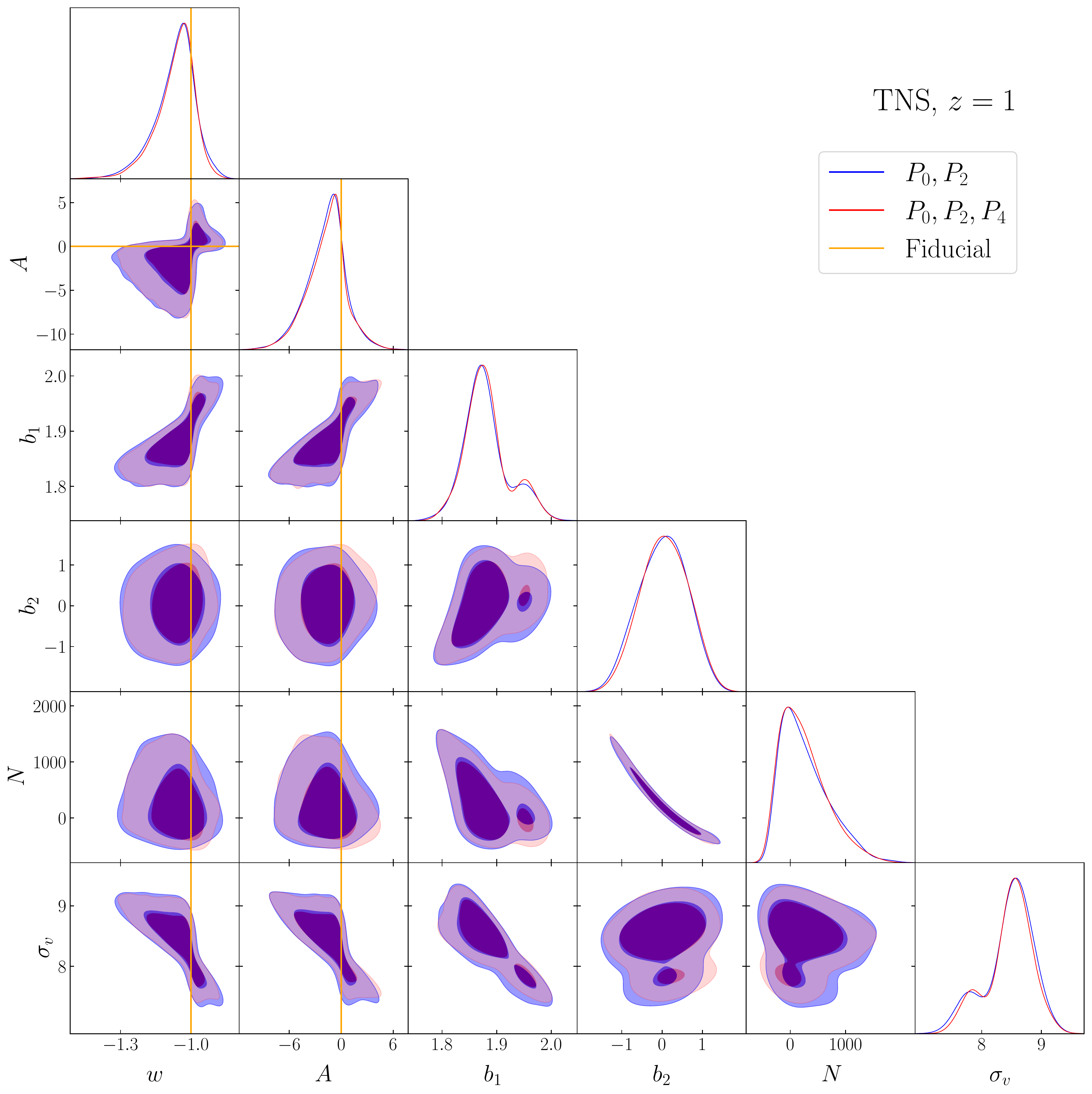}
    \caption{Marginal posterior distributions for all sampled parameters in the TNS model for $z=1$, comparing the case with only the monopole and quadrupole for $k_{\rm max}^{\ell=0,2}=0.22~h/{\rm Mpc}$ with the addition of the hexadecapole with $k_{\rm max}^{\ell=4}=0.05~h/{\rm Mpc}$.}
    \label{MCMC_tns_compP4}
\end{figure}

\begin{figure}[h]
    \centering
		\includegraphics[width=\textwidth]{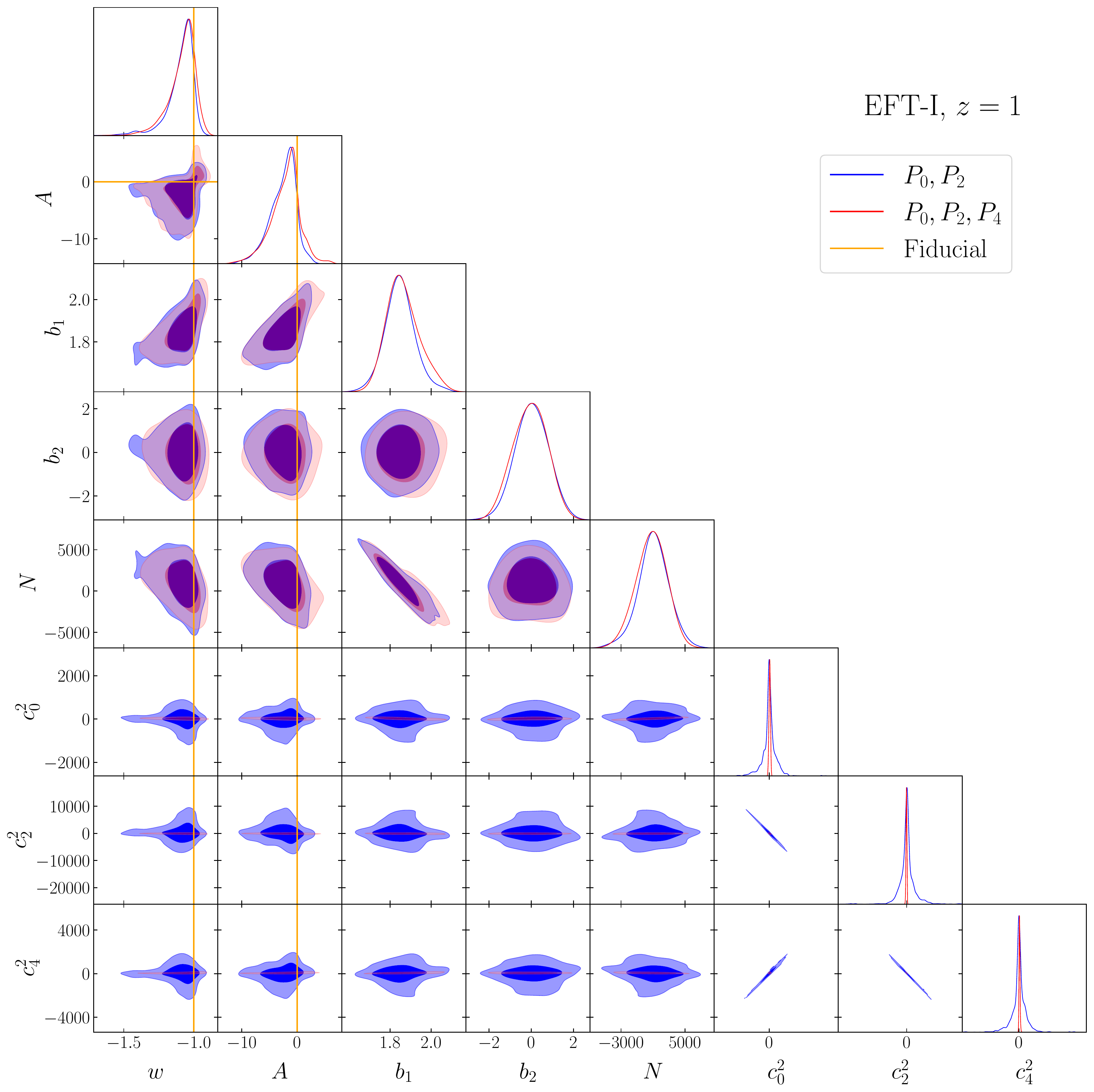}
    \caption{Marginal posterior distributions for all sampled parameters in the EFT-I model for $z=1$, comparing the case with only the monopole and quadrupole for $k_{\rm max}^{\ell=0,2}=0.19~h/{\rm Mpc}$ with the addition of the hexadecapole with $k_{\rm max}^{\ell=4}=0.05~h/{\rm Mpc}$.}
    \label{MCMC_eft_compP4}
\end{figure}

\begin{figure}[h]
    \centering
		\includegraphics[width=\textwidth]{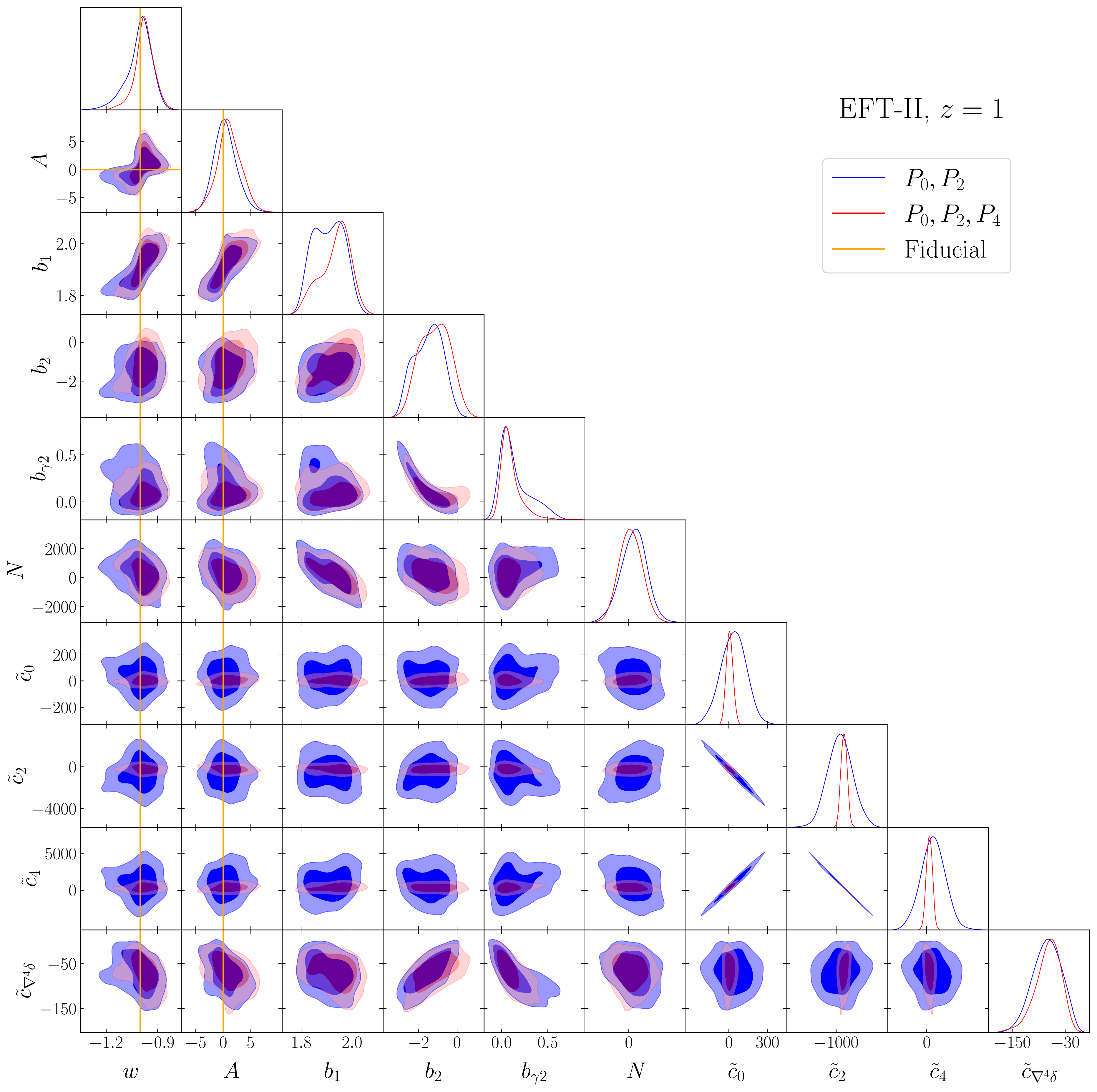}
    \caption{Marginal posterior distributions for all sampled parameters in the EFT-II model for $z=1$, comparing the case with only the monopole and quadrupole for $k_{\rm max}^{\ell=0,2}=0.3~h/{\rm Mpc}$ with the addition of the hexadecapole with $k_{\rm max}^{\ell=4}=0.05~h/{\rm Mpc}$.}
    \label{MCMC_eft2_compP4}
\end{figure}

In this appendix we show full contour plots, including all nuisance parameters for each model. 
In Figs.~\ref{MCMC_tns_compP4},~\ref{MCMC_eft_compP4} and \ref{MCMC_eft2_compP4}, we demonstrate the effect of the hexadecapole for TNS, EFT-I and EFT-II, respectively. For TNS the effect is subtle as we only include scales up to $k_{\rm max}^{\ell=4}=0.05~h/{\rm Mpc}$, but for the EFT-based models this small addition of hexadecapole data is sufficient break the degeneracy between the EFT counterterms. As mentioned in the main text, this is due to the existence of one combination of the three $O(k^2)$ counterterms that can only be measured by the hexadecapole. It is this freedom that causes the EFT-based models to have less biased results when adding hexadecapole data, but also what makes this additional data uninformative in what concerns the other nuisance parameters, as it is only used to measure this combination of counterterms.

\begin{figure}[h]
    \centering
		\includegraphics[width=\textwidth]{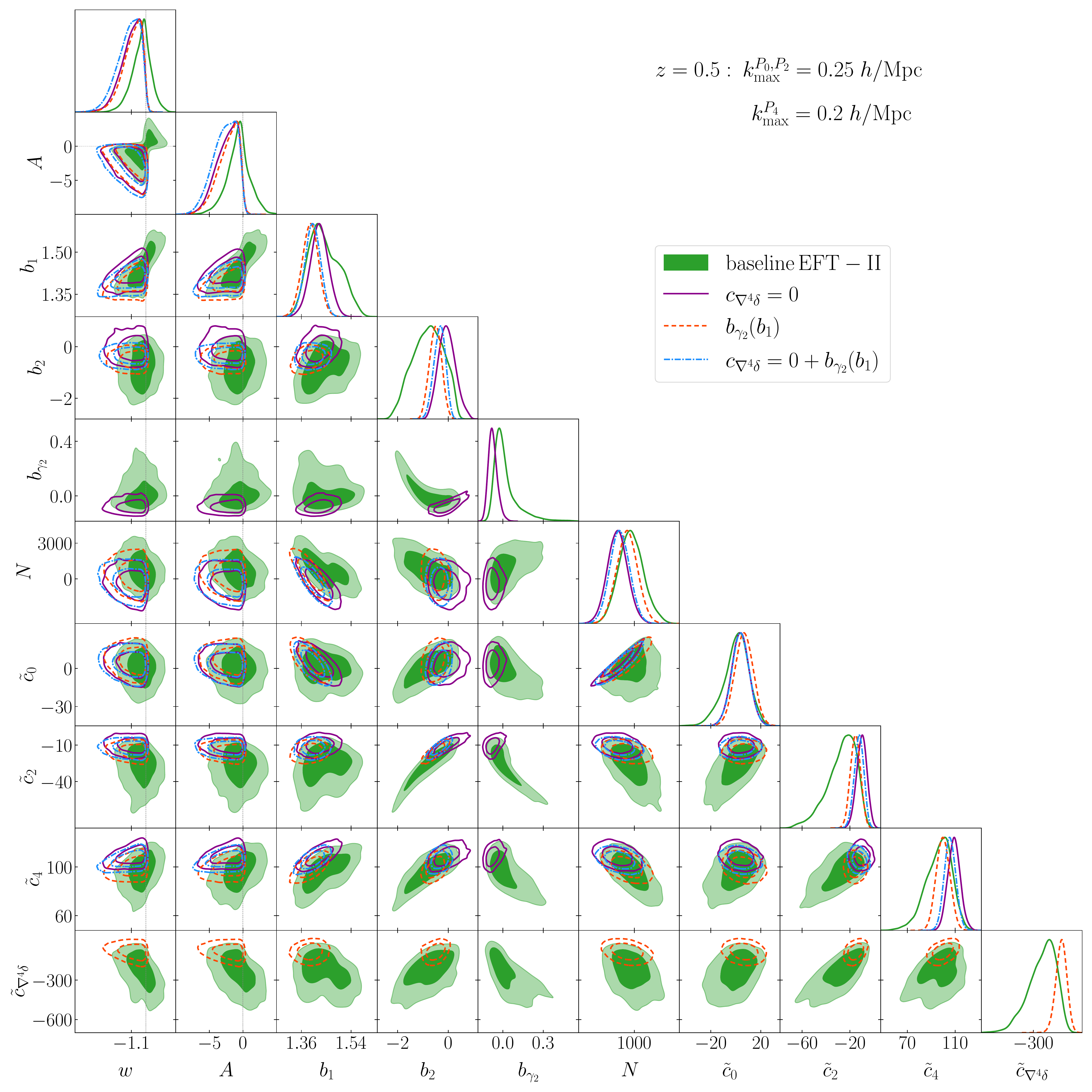}
    \caption{Marginal posterior distributions for all sampled parameters in the tested versions of the EFT-II model for $z=0.5$ with $k_{\rm max}^{\ell=0,2}=0.25~h/{\rm Mpc}$ and $k_{\rm max}^{\ell=4}=0.2~h/{\rm Mpc}$. Filled green contours: baseline EFT-II model, with $\tilde{c}_{\nabla^4 \delta}$ anf $b_{\gamma_2}$ free, solid purple lines: model with $\tilde{c}_{\nabla^4 \delta}=0$, dashed red lines: model with the local-Lagrangian relation $b_{\gamma_2}(b_1)$, dot-dash blue lines: model with $\tilde{c}_{\nabla^4 \delta}=0$ and the local-Lagrangian relation.}
    \label{eft2-diff-bias-full}
\end{figure}

Finally, in Fig.~\ref{eft2-diff-bias-full}, we show the full contours for the different versions of the EFT-II model, used to investigate the differences with EFT-I. As concluded in the main text, this test shows that all free parameters used in the baseline EFT-II model are essential to its success.

\end{document}